# Traveling magnetopause distortion related to a large-scale magnetosheath plasma jet: THEMIS and ground-based observations

A. V. Dmitriev[1,2], A. V. Suvorova[2,3]

**Abstract** Here, we present a case study of THEMIS and ground-based observations on the dayside magnetopause, and geomagnetic field perturbations related to the interaction of an interplanetary directional discontinuity (DD), as observed by ACE, within the magnetosphere on 16 June 2007. The interaction resulted in a large-scale local magnetopause distortion of an "expansion – compression – expansion" (ECE) sequence that lasted for ~15 min. The compression was caused by a very dense, cold, and fast high-$\beta$ magnetosheath plasma flow, a so-called plasma jet, whose kinetic energy was approximately three times higher than the energy of the incident solar wind. The plasma jet resulted in the effective penetration of the magnetosheath plasma inside the magnetosphere. A strong distortion of the Chapman-Ferraro current in the ECE sequence generated a tripolar magnetic pulse "decrease – peak– decrease" (DPD) that was observed at low and middle latitudes by the INTERMAGNET network of ground-based magnetometers. The characteristics of the ECE sequence and the spatial-temporal dynamics of the DPD pulse were found to be very different from any reported patterns of DD interactions within the magnetosphere. The observed features only partially resembled structures such as FTE, hot flow anomalies, and transient density events. Thus, it is difficult to explain them in the context of existing models.

# 1. Introduction

Multi-satellite observations allow detailed studys of localized magnetopause distortions having spatial scales of the order of Earth's radius (Re). In contrast to global magnetosphere disturbances caused by interplanetary shocks and solar wind pressure variations, local distortions are driven by phenomena such as flux transfer events (FTE), Kelvin – Helmholtz (KH) waves, and the magnetosheath pressure pulses related to magnetic reconnections, flow shears, hot flow anomalies (HFAs), foreshock cavities, and transient density events (TDEs). The various phenomena have similar features that make it difficult to determine the causes of observed magnetosphere disturbances.

Dayside magnetopause distortions are often associated with FTEs. The main general feature is a characteristic bipolar signature in the $B_n$ magnetic field component, which is normal to the nominal plane of the magnetopause [e.g. *Elphic*, 1995]. Standard +/- FTE $B_n$ signatures tend to be found north of the ecliptic plane, and reverse -/+ signatures south of the plane [*Berchem and Russell*, 1984; *Kawano et al.*, 1992]. FTE magnetic effects are much larger than variations that the magnetosheath pressure could produce. Characteristic spatial and temporal scales of the FTE are ~1 Re and ~5 min, respectively. FTEs have such a large latitudinal extension that the same FTE can be observed simultaneously in the Northern and Southern Hemispheres [*Elphic and Southwood*, 1987]. Multi-satellite observations show a specific spatial pattern for plasma and magnetic field dynamics related to the FTE [*Hasegawa et al.*, 2006; *Sibeck et al.*, 2008]. Roughly, flux ropes of spiraling magnetic fields, generated during reconnections, are embedded within a broadened current layer of weak magnetic field strengths with mixtures of magnetosheath and magnetospheric plasmas.

Strong magnetopause distortions in the flank and tail regions are related to the KH instability [e.g. *Fairfield et al.*, 2000, and references therein]. The KH instability at the magnetopause results in large amplitude waves. Recently, *Eriksson et al.* [2009] reported a post-noon event for large-scale magnetosheath flow vortices of KH waves intermittent with magnetic islands of FTEs.

HFAs are characterized by greatly heated low-density solar wind plasmas with a depleted magnetic field and a substantial flow deflection flanked by regions of enhanced plasma density and magnetic field strength [*Thomsen et al.*, 1986; *Sibeck et al.*, 2003], and have a spatial scale of ~1 to 3 Re. HFAs are generated in the foreshock region during interactions of a solar wind tangential discontinuity (TD) with the quasi-parallel (hereafter $Q_\parallel$) side of the bow shock [*Lin*, 2002; *Omidi and Sibeck*, 2007; *Facskó et al.*, 2009]. HFA signatures are transmitted from the foreshock to the magnetosheath. Increased magnetosheath density causes a local magnetopause compression with a depth of a few Re. The depressed pressure briefly allows the magnetopause to move outward at a distance of several Re, far from the undisturbed location [*Sibeck et al.*, 1998; *Jacobsen et al.*, 2009]. Fast indentations and expansions of the magnetopause result in the substantial deviation of the geomagnetic field and the appearance of magnetic signatures similar to FTE, such as the bipolar signature in the $B_n$.

Foreshock cavities are similar to HFAs [*Sibeck et al.*, 2002; *Turner et al.*, 2011] and have durations of several minutes, reduced densities, and magnetic field strengths flanked by enhancements and enhanced ion temperatures. However, within cavities there is no great reduction or deflection of flow velocities, the ion populations are always anisotropic, and the ion temperatures are lower than those inside HFAs. Foreshock diamagnetic cavities are generated in turbulent foreshocks due to the interaction of the oncoming solar wind with a beam of ions backstreaming from the $Q_\parallel$ bow shock [*Thomas and Brecht*, 1988; *Lin and Wang*, 2005]. Magnetopause disturbances related to foreshock cavities are similar to those in HFAs.

Transient density events are magnetosheath structures characterized by an enhanced plasma density and a depressed magnetic field [*Hubert and Harvey*, 2000]. The observed thickness of this structure is approximately 0.5 $R_E$ along the GSE X axis, and is convected from the bow shock to the magnetopause without spreading. The interaction of TDE with the magnetopause shows that plasma can impulsively penetrate into the dayside magnetosphere. The TDE is generated by interactions between the bow shock and an interplanetary rotational discontinuity (RD) [*Lin et al.*, 1996; *Cable and Lin*, 1998; *Tsubouchi and Matsumoto*, 2005]. MHD and hybrid simulations predict that interactions between the bow shock and an RD results in the generation of upstream and downstream pressure pulses propagating anti-sunward. The downstream pulse is generated in the magnetosheath by a structure that consists of an RD and two slow shocks. The amplitude of these pressure pulses can be up to 100% of the background magnetosheath value. A substantial increase in pressure is mainly caused by the deflection of the plasma flow tangential to the bow shock's normal direction. The upstream pressure pulse is generated in the foreshock region by reflected ions when a local $Q_\parallel$ bow shock becomes a quasi-perpendicular (hereafter $Q_\perp$) shock. An upstream pulse convects through and interacts with the bow shock, producing a pressure pulse in the downstream region. The pressure pulse quickly propagates with magnetosonic speed and can catch up to the slowly propagating downstream pulse.

In a number of cases, the kinetic energy of magnetosheath pressure pulses exceeds the background level [*Němeček et al.*, 1998; *Savin et al.*, 2004, 2008]. Such so-called transient flux enhancements (TFEs), or magnetosheath plasma jets, are characterized by intense localized fast ion fluxes, whose velocities vary from 100 to 600 km/s, and whose kinetic energy density can be two to five times higher than that in the upstream solar wind. The dimensions of jets are on the order of ~1 Re, and the duration varies from tens of seconds to a few minutes. *Savin et al.* [2008] presented observations of high kinetic energy density jets based on Interball and Cluster data, and argued that the presence of these jets was due to non-fully developed turbulence in the magnetosheath that remains intermittent and structured in time and space. The authors suggested that such jets pierce through the magnetopause and, therefore, could provide the respective diffusion-like plasma transport toward the magnetosphere. *Amata et al.* [2006] and *Savin et al.* [2006] reported experimental support for this mechanism. From Cluster data, they found a kinetic effect for the penetration of magnetosheath ions with a finite gyroradius (by protons with an energy > 350 eV) through a thin current sheet at the magnetopause (~90 km thickness).

On the other hand, *Sibeck et al.* [2000] found that in the near-tail region TFEs can relate to HFAs, and that they can result in magnetopause undulations and prominent geomagnetic variations. Recently, *Amata et al.* [2011] reported Cluster observations for a deep indentation in the high-latitude magnetopause caused by an anti-sunward jet ramming into the magnetopause. However, there have been no detailed studies based on direct observations of jet interactions with the magnetopause.

Variations in magnetosheath pressure alternately compress and rarefy the dayside magnetospheric magnetic field by launching MHD (compressional and Alfvén) waves that propagate through the magnetosphere to the ionosphere [e.g. *Chi et al.*, 2006]. The propagation is quite fast, but non-uniform, as a result of the dipole configuration of the geomagnetic field. At low and middle geomagnetic latitudes below the plasmapause latitude of ~60°, a signal from the magnetopause propagates through the dense plasmasphere relatively slowly (~1.5 min). *Sibeck et al.* [2003] reported HFA-related magnetic signatures of ~20 min in duration that were detected in a geosynchronous orbit and with low-latitude ground magnetometers. At higher latitudes above the plasmapause, magnetopause compression is mainly transmitted by Alfvén waves propagating very quickly (< 1 min) along the magnetic field lines to the ionosphere. In this region, transient compression of the magnetopause results in a pair of oppositely directed vortices of ionospheric currents, the so-called traveling convection vortices observed by high-latitude magnetometers [*Friis-Christensen et al.*, 1988; *Sibeck et al.*, 2003; *Jacobsen et al.*, 2009]. Hence, ground magnetic variations can be used for the analysis of magnetopause distortions.

In the present paper, we report a case event for dayside magnetopause distortions and geomagnetic disturbances related to the interaction of an interplanetary discontinuity with the magnetosphere on 16 June 2007. Observations of the magnetosheath jet and magnetopause dynamics, performed using a Time History of Events and Macroscale Interactions during Substorm (THEMIS) multi-spacecraft experiments, are presented and analyzed in Section 2. Upstream solar wind conditions are described in Section 3. Geomagnetic responses to magnetopause distortions were studied using the INTERMAGNET network of ground-based magnetometers in Section 4. Results are discussed and summarized in Section 5.

## 2. Magnetopause distortions and the magnetosheath jet

THEMIS observations of the plasma and magnetic fields at 0440 – 0510 UT on June 16 2007 are shown in Figure 1. The chosen interval was quiet, as follows: geomagnetic indices $Dst$~0 and $AE$~100 nT. The magnetic field was measured with a time resolution of ~ 3 sec using the THEMIS/FGM instrument [*Auster et al.*, 2008]. Plasma was measured with THEMIS/ESA instruments [*McFadden et al.*, 2008]. During the time interval considered, high-resolution (~3 sec) plasma spectra were only available for the THE/ESA and THB/ESA instruments that operated in fast survey mode. Note that these high-resolution data with a reduced distribution were highly consistent with the full distribution data, that had a lower time resolution of ~1.5 min (not shown).

The GSM coordinates of THEMIS probes are presented in Table 1. THEMIS is located in the postnoon region and slightly southward of the GSM equator. As shown in Figure 1, all of the THEMIS probes, with the exception of THA, were, most of time, located within the magnetosphere, a region with a smoothly varying strong magnetic field with a magnitude of ~40 nT and a low-density hot (> 10 keV) plasma. The outermost THA probe was mainly located in the magnetosheath, a region with a highly variable magnetic field. During this period of time, other main near-Earth satellites could not provide useful data. Geotail and GOES were located on the night side of the magnetosphere, the Double Star satellites were in the inner magnetosphere, and Cluster was in the high-latitude magnetospheric region.

Variations of the $By$ and $Bz$ components of the magnetic field measured by THA in the magnetosheath correlated well with the $Bz$ and $By$ components of the interplanetary magnetic field (IMF) observed 46.5 min earlier by the ACE upstream monitor near the L1 point. We obtained correlation coefficients of $r$ = 0.6 and 0.7, respectively, for $By$ and $Bz$ variations during magnetosheath intervals at 0440 - 0447 UT, 0453 - 0457 UT, and 0501 – 0510 UT. The delayed IMF $By$ and $Bz$ components changed signs between 0447 and 0453 UT. During this time, the THA probe entered the magnetosphere and magnetic measurements in the magnetosheath were not available.

During the THA, magnetosphere entries from 0447 to 0453 UT and from 0459 to 0501 UT, other probes observed a depletion of the geomagnetic field from 0446 to 0452 UT and from 0456 to 0502 UT. The amplitude of the depletion measured by THE was ~15 nT and ~5 nT, respectively, at ~0449 and ~0459 UT. Such dynamics can be interpreted as magnetopause expansions. Between expansions, THEMIS observed a strong magnetopause compression accompanied by a significant enhancement of the geomagnetic field from ~45 to 60 nT, as seen by THB. The magnetopause moved inward such that the probes THC, THD, and THE moved from ~0454 to ~0456 UT in the magnetosheath, a region of dense and cold (< 1 keV) plasma streams. Therefore, from 0447 to 0502 UT THEMIS observed a magnetopause distortion of an "expansion – compression – expansion" (ECE) sequence.

Here, it is important to point out that the magnetopause compression was accompanied by a penetration of magnetosheath plasma inside the magnetosphere. From 04:54:30 UT, the innermost THB probe in the magnetosphere detected ions and electrons with mixed spectra. Together with magnetospheric high-energy ions and thermal electrons, the magnetosheath population of cold ions and hot (> 100 eV) electrons can be observed. Following the end of compression at ~0457 UT the population of magnetosheath electrons diminished. In contrast, prominent fluxes of magnetosheath ions persisted in the magnetosphere until ~0508 UT (i.e. during more than 10 min). Note that the most intense splashes in the magnetosheath plasma in the magnetosphere were detected by THB from 04:55:30 to ~04:57:30 UT (i.e. during the decompression).

We can assume that magnetosheath plasma penetration, inside the magnetosphere, becomes possible due to substantial distortions in the magnetopause shape. In order to study the distortions, we determined the magnetopause orientation (i.e. normals to the magnetopause). Under the northward IMF, the low- and mid-latitude magnetopause is a tangential discontinuity (TD) that separates the magnetospheric plasma and the magnetic field from the magnetosheath. During reconnection under a southward IMF, the magnetopause can be presented as a rotational discontinuity (RD). In Figure 1, one can clearly see a strong rotation in the magnetic field at the magnetopause that is proper for the TD. On the other hand, the negative $B_z$ and the presence of the magnetosheath plasma in the magnetosphere indicates that the magnetopause may be an RD. We applied a Walén test [*Sonnerup et al.*, 1987] for THE magnetopause crossings at ~04:54:20 and 04:56:40 UT for which high-resolution plasma data were available. We found a very small slope of $S < 0.1$ and a correlation of $r < 0.2$ for the scatter plot of $| V - V_{th} |$ versus $V_a$, where $V$ is a vector of the plasma velocity, $V_{th}$ is the vector velocity of the deHoffmann-Teller frame, and $V_a$ is the Alfvén speed. The failure of the Walén test indicates that the magnetopause is TD rather than RD.

Using a minimal variance analysis (MVA), we calculated a normal to the plane of magnetic field rotation (i.e. a normal to the magnetopause). Table 2 presents the time intervals used for the MVA, the location of magnetopause crossings, the eigenvectors, and the eigenvalues of the normals. By taking into account time delays between magnetopause crossings, we estimated that the magnetopause moves inward (at ~0454 UT) and outward (at ~0456 UT) with the speed of ~80 ± 10 km/s and ~200 ± 50 km/s, respectively. The speed was calculated along the normal to *Lins et al.'s* [2010] reference magnetopause of **n** = (GSM 0.73, 0.66, -0.17). We used the solar wind conditions observed by ACE at 0355 UT (see Figure 3). Note that the upstream conditions did not change very much. Therefore, the reference magnetopause remained almost constant for the entire interval considered.

Here, we emphasize that the observed magnetopause orientation (see Table 2) deviated substantially from the model. Therefore, actual magnetopause speed may be different. We estimated the actual speed of inward and outward magnetopause motion, observed by THE, THC, and THD respectively, at 0454 and ~0456 UT. By taking into account the location of THE and THD and by calculating the average eigenvectors obtained for THE, THC, and THD magnetopause crossings, we determined an inward and outward magnetopause speed of ~73 km/s and 112 km/s, respectively. For THE, THC, and THD magnetopause crossings at 0454 UT and 0456 UT, we found that the magnetic field rotation at the magnetopause lasted for ~ 9 ± 3 and ~ 6 ± 3 sec, respectively. Note that for the latter case, the magnetic field was perturbed and, thus, it was difficult to accurately determine the duration of its rotation. Magnetopause thickness was estimated to be approximately 650 km, in good agreement with the average equatorial value of ~500 km [*Berchem and Russell*, 1982].

For a further numerical analysis of magnetopause distortions (see Figure 2), we converted the vectors of the THEMIS magnetic field and the plasma velocity, as well as magnetopause normals, into normal coordinates (**l**, **m**, and **n**) in the frame of *Lin et al.'s* [2010] reference magnetopause. For the reference magnetopause at the THEMIS location, we calculated a distance of ~12.2 Re; **l** = (GSM 0.09, 0.15, 0.98) is in the magnetopause plane and points northward in the direction of the magnetospheric magnetic field; **n** = (GSM 0.73, 0.66, -0.17) is the magnetopause normal that points outward; and **m** = (GSM 0.67, -0.74, 0.05) completes the triad by pointing dawnward. For convenience, we used an opposite component (i.e. –**m** pointing duskward similar to the Y-GSM axis). In the third panel from the top in Figure 2, magnetopause normals are shown in planes **l-n** (meridional cut) and **m-n** (zonal cut). A sketch of the magnetopause location was drawn using a dashed line smoothly connecting the magnetopause crossings. From the location of the crossings, we determined that the magnetopause displacement varied from -0.7 Re to 0.7 Re. Hence, the amplitude of the magnetopause distortion was found to be at least ~ 1.4 Re.

Using THE data (the two upper panels in Figure 2), we found that the magnetosphere compression at ~0454 related to a very dense (~20 cm$^{-3}$), cold (~0.2 keV), and fast magnetosheath plasma flow (~300 km/s). We calculated the total magnetosheath pressure, *P*tot, as a sum of the magnetic (*Pm*), the thermal (*Pt*), and the dynamic (*Pd*) pressures. The thermal pressure (*Pt*) is a sum of ion and electron thermal pressures. One can see that in the compression region, the magnetosheath thermal pressure is ~1 nPa and exceeds the magnetic pressure of ~0.5 nPa. Note that a very low magnetosheath magnetic field was also observed by THA from 04:54:30 to 04:56:00 UT. In the maximum of *Ptot* from 04:54:20 to 04:54:50 UT, the sum of *Pt* and *Pm* (~1.6 nPa) was close to the total pressure of the solar wind of *Psw*~1.7 nPa as calculated from ACE measurements (see Figure 3). In addition to thermal and magnetic pressures, the magnetosheath plasma has a high speed and, hence, a high kinetic energy - *Pd* ~ 3 nPa which resulted in very strong total magnetosheath pressure of *Ptot* ~ 4.4 nPa that was more than two times higher than the solar wind pressure - *Psw*. Therefore, we can interpret this high-energy magnetosheath structure as a high-beta plasma jet [*Nemecek et al.*, 1998; *Savin et al.*, 2008].

The jet was preceded by a gradual enhancement of *Ptot* beginning at 0452 UT. The *Ptot* enhancement and the corresponding inward magnetopause motion was manifest as an enhanced magnitude in the geomagnetic field accompanied by an increasing inward drift ($V_n < 0$) of the magnetospheric plasma. The innermost THB probe (see the bottom panel) detected a strong positive pulse of ~15 nT in the tangential magnetic field components $B_l$ and $B_m$, peaking at 0455 UT that corresponded to a strong magnetopause compression produced by the jet.

Variations in the plasma streams observed in the magnetosheath and magnetosphere by THE and THB are shown, respectively, in the fourth and fifth panels of Figure 2. Within the jet and in close vicinity to the magnetopause from 04:54:24 to 04:55:30 UT, THE observed fast magnetosheath plasma streams of 250 to 300 km/s directed duskward (-$V_m$ ~ 200 km/s) and southward ($V_l$ ~ -150 km/s). Note that the southward stream appears to be too fast for the THEMIS location slightly southward from the GSM equator (GSM lon ~ 50° lat ~ -15°) where duskward streams should be dominant. Here, we need to point out that fast magnetosheath streams have a strong inward component of $V_n$ ~ -100 km/s. At the same time, inward plasma flows with a $V_n$ ~ -50 km/s were also observed by the THB inside the magnetosphere, in spite of the fact that the magnetopause began to move outward at 0455 UT. Observed plasma dynamics indicated an intense interaction of the plasma jet with the magnetopause that could result in a distortion of the magnetopause shape.

A jet-related magnetopause compression was revealed by THE, THC, and THD as a brief magnetosheath encounter from ~0454 to ~0456 UT (see the third panel from the top in Figure 2). The normals listed in Table 2 suggest that the local magnetopause undergoes an indentation. Important to note is that during this time, the magnetopause is not in balance, as follows: the total magnetospheric pressure (*P*tot) is substantially smaller than that in the adjacent magnetosheath. Therefore, magnetopause dynamics are in non-equilibrium and transient. The magnetopause normal at 0454 – 0455 UT is tilted essentially duskward and slightly southward relative to the nominal one. The orientation of the magnetopause corresponds to the incident plasma stream flowing tangentially to the distorted magnetopause toward southern-dusk and inward in accordance with THE plasma observations. Therefore, it is reasonable to suggest that abnormal southward and inward plasma streams are produced by the jet.

From ~04:55:30 to 04:56:20 UT, the magnetosheath plasma velocity and density decreased such that the magnetosheath pressure (*P*tot) was equal to the solar wind pressure (*Psw*). Magnetosheath plasma streams are almost parallel to the nominal magnetopause (i.e. $V_n$ = 0 km/s at THE). In the magnetosphere, THB observes sporadic streams of the mixed plasma (see Figure 1) directed duskward and southward with a quite large $V_n$ ~ -100 km/s. Perhaps these streams are contributed by the plasma that penetrates from the magnetosheath.

The interval from 04:56:20 to 04:56:37 UT is interesting to consider when intense streams of the dense magnetosheath plasma are detected in the magnetosphere by the THB. During this time interval, THE observed magnetosheath plasma streams adjacent to the magnetopause. The streams were directed mainly duskward (-$V_m$ ~ 200 km/s) and slightly inward ($V_n$ ~ -50 km/s). Important to note is that the meridional component of the plasma velocity $V_l$ turned into zero. In this region it seems that the magnetosheath plasma stream was deflected toward dusk, an assumption that is supported by the magnetopause orientation observed during outbound crossings at ~04:56:37 UT. The magnetopause normal was tilted essentially duskward and northward from the undisturbed direction that corresponded to the observed magnetosheath streams tangential to the distorted magnetopause.

Fast magnetopause expansion was observed from 04:56:37 to 04:57:30 UT, when both the THE and THB detected outward streams of magnetospheric plasma with speeds of $V_n$ ~100 to 300 km/s. The speeds are consistent with previous estimations of magnetopause outward motion with velocities of ~200 km/s. The fast magnetopause expansion was accompanied by a growth of the magnetosheath plasma population in the magnetosphere as observed by THB (See Figure 1). Therefore, the most intense penetration of the magnetosheath plasma into the magnetosphere occurred during magnetopause decompression immediately following the jet passage.

As a result of the relatively low speed of THEMIS probes, the temporal dynamics of the magnetic field and the plasma observed by THEMIS between ~0453 and 0457 UT resulted from the propagation of the indentation along the magnetopause. The deviation of the magnetopause normal in the **l-n** (meridional) plane changed sequentially from southward to northward. The dynamic means that the indentation moved toward the north. In the **m-n** (zonal) plane, the magnetopause normal was tilted duskward and practically did not change, making it difficult to determine the propagation direction.

The indentation was preceded and followed by magnetopause expansions that were accompanied by corresponding variations in the plasma and magnetic fields observed in the magnetosphere by THEMIS. Strong deviations of the magnetopause normals during the preceding expansion from 0446 to 0452 UT suggest a bulge-like magnetopause expansion. The temporal dynamics of the normal tilt, changing from northern-dusk to southern-dawn, corresponded to the propagation of the bulge from southern-dawn to northern-dusk.

The following expansion, lasting from 0457 to 0501 UT, was characterized by fast variations in the magnetopause as observed by THA and by the complex dynamics of the magnetospheric plasma as observed by THE. Under such conditions, the MVA failed and, hence, the magnetopause normals could not be determined. As a result, it was difficult to analyze the geometry of the expansion outlined below.

## 3. Upstream conditions and directional discontinuity

In the previous section, we estimated a time lag of 46.5 min between the ACE upstream monitor and the THA probe. ACE observations of the solar wind plasma and the IMF in the time interval from 0345 to 0415 UT on 16 June 2007 are shown in Figure 3. For this time interval, the GSM coordinates of ACE are the following: $X \sim 231.6$ Re, $Y \sim -37.7$ Re, and $Z \sim 12.7$ Re. The magnetic field and the plasma parameters were measured with a 1-sec and a 1-min resolution, respectively. Interplanetary conditions were quiet and varied slightly. The total solar wind pressure ($P_{sw}$) (i.e., the sum of the solar wind dynamics, and the thermal and magnetic pressures) varied between 1.6 and 2 nPa. At ~0401 UT, ACE observed a directional discontinuity (DD). Across the DD, the $B$y and $B$z, and $V$y and $V$z components overturned while the IMF magnitude, as well as the plasma characteristics did not (practically) change. Note that the Wind satellite during that time was located very far from the Sun-Earth line and observed a different solar wind structure.

When the DD slides along the dayside magnetosphere the local cone angle $\theta_{Bn}$ between the IMF vector and the normal to the bow shock changes dramatically. Figure 4 shows the contours of the equal cone angles (iso-contours) calculated using *Chao et al.'s* [2002] bow shock model for three different time moments, as follows: at 0355 UT (i.e. before DD arrival); in the center of the DD at ~04:00:46 UT; and at 0406 UT (after DD passing). Prior to DD arrival, the quasi-parallel ($\theta_{Bn} < 25°$) portion of the bow shock was located in the postnoon - dusk sector of the Southern Hemisphere. Then, the portion travelled northward and dawnward, passing the subsolar region and reaching its post-DD equilibrium position in the prenoon - dawn sector of the Northern Hemisphere. In Figure 4, a transitions corridor for the bow-shock portion with a $\theta_{Bn} < 25°$ is restricted by dotted lines. The corridor is based on calculations of the $\theta_{Bn}$ iso-contours for various time moments between 0355 and 0404 UT.

During the transition, the $Q_{\parallel}$ portion is replaced by $Q_{\perp}$ in the south-dusk sector, and the $Q_{\perp}$ - $Q_{\parallel}$ $Q_{\perp}$ transition occurs in a ~40° area surrounding the subsolar point. Such variations of the $\theta_{Bn}$ cause dramatic changes in bow shock formation [e.g. *Russell and Petrinec*, 1996; *Eastwood et al.*, 2005].

Hence, the interaction of DD with the bow shock may result in magnetospheric disturbances. However, this interaction is a transition, as demonstrated in Figure 4b by dashed iso-contours. As a result of the transition at a given time moment, the region of small $\theta_{Bn}$ occupies only a portion of the subsolar region, while another portion is already converted to the $Q_{\perp}$. The temporal and spatial sequence of the bow shock transition depends on the DD orientation.

The orientation of DD is determined by the method of MVA. Note that the results of MVA vary by the length of the time interval chosen for the analysis. If the results vary greatly for different time intervals, they are unreliable. By varying the boundaries of the time intervals, we found time ranges for which the results of MVA changed slightly and gradually (i.e. the solution of MVA was stable). The left and right boundaries allocated the time ranges, respectively, at 03:59:10 – 04:00:10 UT and at 04:01:10 – 04:02:05 UT, which were free from interplanetary discontinuities. For this choice, we obtained, in GSM coordinates, the normal **n** = (-0.25 ± 0.01, 0.70 ± 0.01, 0.67 ± 0.01) with the corresponding eigenvalues (3.3 ± 0.1, 0.09 ± 0.01, 0.02 ± 0.003).

We also estimated a component of the magnetic field normal to the DD of approximately $B_n = 0.7$ nT. The ratio between $B_n$ and the total magnetic field $|B|$, averaged over the time range of the discontinuity (from 0357 to 0402 UT), was estimated to be equal to $B_n/|B| \sim 0.2$. Since $B_n$ was relatively small, we could normally obtain DD's independently by using a cross-product method [e.g. *Lin et al.*, 2009]. Before the discontinuity from 0347 to 0359 UT and afterward from 04:01:10 to 04:02:05 UT, we found average vectors of IMF in GSM, respectively, of $B_1 = (-2.38, -2.02, 1.41)$ nT and $B_2 = (-2.08, 1.16, -2.01)$ nT. Vector multiplication gave a normal to the IMF rotation plane **n'** = (-0.2, 0.67, 0.71). The normal was quite close to that obtained by the method of MVA. The difference may be due to the finite value of the $B_n$. Hence, we could consider the result of MVAs as a reasonable approach to the normal DD's.

The propagation time of DDs from ACE to the magnetosphere depends on the type of DD. TDs are frozen and move with the solar wind. An RD propagates in the solar wind frame with an Alfvén velocity ($V_a$) parallel or anti-parallel to the normal **n**. A small ratio of $B_n/|B|$ indicates a TD [e.g. *Neugebauer et* *al.*, 1984]. On the other hand, the ratio of the magnetic field magnitude changing across the DD ($\Delta B$) to $|B|$ is also very small, $\Delta B/|B| < 0.1$, and proper for RD. Our efforts to apply a Walén test in the range from 0357 to 0402 UT yielded ambiguous results. We found the velocity of the deHoffmann-Teller frame in GSM: $V_{th}$ = (-505, 20, 22) km/s. However, for the scatter plot of $|V - V_{th}|$ versus $V$a, we obtained a relatively small slope, $S \sim -0.46$, with a correlation of $r \sim 0.84$. As a result of the low temporal resolution (1 min) for ACE plasma data, we were unable to perform the Walén test accurately. Hence, the available data did not allow the unambiguous determination of the DD type. However, a number of recent studies demonstrate that DD, with a very small but non-zero $B_n$ and a relatively small slope $S$, can be attributed to the RD [e.g. *Lin et al.*, 2009; *Teh et al.*, 2011].

For given solar wind plasma conditions and $B_n \sim 0.7$ nT, we obtained $V$a $\sim 10$ km/s along the normal of the DD. By taking into account the solar wind velocity (GSM $V$x = -500, $V$y = 30, $V$z = 13 km/s), the $V$a, the normal **n**, and the ACE location, we could estimate the DD propagation time (d$T$) to the bow shock. Here, we provide the RD type of discontinuity propagating anti-sunward with a planar geometry. Figure 5 shows iso-contours of d$T$ for the bow shock calculated from the *Chao et al.* [2002] model. Note that the propagation time calculated for TD is ~2 min larger.

The orientation of DD is such that, on reaching the bow shock, the first hit should be on the southward-dawnward portion at ~0435 UT. The DD would then slide across the subsolar region at ~0447 UT toward northern-dusk. Moving across the subsolar region, the DD passes the transition corridor within ~12 min from ~0443 UT to ~0455 UT (d$T$ increases from ~42 min to ~54 min). During this time interval, the quasi-parallel portion of the bow shock is translated from south-dusk to north-dawn and is replaced by Q$_\perp$.

Note that the propagation time (d$T$) and the width of the transition corridor are determined within an uncertainty of a few minutes. As a result of the large distance to ACE, tilted interplanetary fronts can change in orientation with time [*Weimer et al.*, 2002]. Additionally, the propagation of the tilted DD is strongly impacted by non-radial components of the solar wind, as well as by the velocity of the DD in the solar wind frame. Even small uncertainties of several km/s result in substantial variations (minutes) in the propagation time. Nevertheless, despite the several-minute uncertainty in the temporal sequence, the spatial pattern of the DD interaction with the bow shock is firm.

## 4. Ground-based magnetic variations

The spatial dynamics of magnetopause distortions were studied using 1-min magnetic data provided by an INTERMAGNET network of ground-based magnetometers. For our case, variations of solar wind pressure (*Psw*) were gradual and small (see Figure 3) and the geomagnetic activity was low. Hence, variations in the geomagnetic field (if any) should result from local variations in magnetosheath pressure.

In order to eliminate the effect of the non-uniform propagation of MHD waves in the magnetosphere, we only used magnetic stations located at geomagnetic latitudes below the plasmapause (~60°). In this region, the difference in the propagation time at different latitudes was less than a minute, as shown by *Chi et al.* [2006]. Hence, the effect could be smoothed by using the magnetic data of a 1-min resolution. In this manner, we aimed to retrieve geomagnetic variations related to localized magnetopause current distortions using durations of several minutes.

Figure 5 shows the location of INTERMAGNET stations in GSM coordinates, calculated for 0455 UT on 16 June 2007. We only utilized magnetic stations located at low and middle geomagnetic latitudes in the dayside, dawn, and dusk sectors (Table 3). We analyzed the temporal variations (d$H$) of the horizontal component ($H$) of geomagnetic field. The variation d$H$ is a module of the vector subtraction of a linear trend ($H_t$), interpolated between points at 0400 and 0530 UT, from the observed vector $H$: $dH = \left| \vec{H} - \vec{H}_t \right|$. Note that the magnitude of d$H$ was always positive and was sensitive to changes in the external electric currents impacting both the strength and the orientation of the magnetic field [*Dmitriev and Yeh*, 2008].

Figure 6 presents temporal variations of d$H$ observed from 0440 to 0510 UT. We determined two types of variations. Figure 6a shows magnetic records characterized by prominent variations of d$H$, with amplitudes of several nT and/or by a well-distinguished magnetic hump around 0455 UT. Magnetic records with relatively weak and irregular variations of d$H$ are presented in Figure 6b.

First, we needed to pay attention to d$H$ variations at equatorial station GUA, whose GSM location was mostly close to THEMIS (i.e. they were radially conjugated). Both GUA and THB observed a very similar (with ~1 min accuracy) pattern of magnetic variations (see Figure 6a), as follows: a tripolar pulse "decrease – peak – decrease" (DPD) with the peak at ~0455 UT and a minima of leading and trailing decreasing at 0450 UT and 0500 UT, respectively. In Figure 5, the peak time (UT) is indicated by four-digit numbers. The presence of the leading (trailing) decrease is indicated by letter "d" before (after) the numbers. The absence of a prominent d$H$ variation is denoted by "####".

The total duration of the DPD pulse at GUA is ~15 min, consistent with THEMIS observations. The amplitude in d$H$ variations of several nT was much smaller than the 15 nT measured by THEMIS and can be explained by the loss of MHD wave energy during propagation through the magnetosphere. Hence, the peak of d$H$ observed at GUA could relate to the localized magnetopause compression caused by the magnetosheath plasma jet, which in turn may result from the DD sliding along the bow shock. The coincidence of magnetic pulse observations in radially conjugated points at ground (GUA) and near the magnetopause (THB) indicated a very short time of ~1 min or less required for the propagation of the signal from the magnetopause to the ground. The fact supports an idea regarding the radial transfer of magnetic signals by fast MHD waves in the magnetosphere [e.g. *Samsonov et al.*, 2011].

In the subsolar region, stations GZH and PHU (Figure 6a) observed almost the same magnetic variations as those at GUA. However, the leading decrease at these stations is not so prominent. The low-latitude station LZH also detected a peak of d$H$ at 0455 UT, but its leading and trailing sides were perturbed by fast magnetic pulsations. Note that fast pulsations of large-amplitudes perturbed the d$H$ variations at stations AAE, AAA, and IRT (indicated by dashed lines) making interpretations of these data difficult.

Slightly different geomagnetic variations were observed at the mid-latitude stations KDU, CTA, ASP, and CNB located southward from GUA. These stations detected no prominent leading decrease and a wide hump of d$H$ lasting from ~0451 to 0457 UT, with a maximum between ~0452 and 0455 UT. The trailing decrease was perturbed by fast magnetic pulsations. The earliest observations of the pulse were found in the southern-dusk sector at stations CNB and EYR. The enhancement of d$H$ at EYR began at ~0450. The maximum represented a wide plateau lasting from 0452 to 0458 UT. For long-lasting enhancements, we used the beginning time as an indicator of magnetic disturbance arrival. The choice seems reasonable since a prolonged magnetic hump may relate to the evolution of a source of geomagnetic disturbance. For a short peak, the beginning of the enhancement almost coincides (within 1 to 2 minutes) with the maximum.

The most prominent DPD pulse, between 0455 and 0457 UT, was revealed at the northern mid-latitude stations. The dayside stations KNY, KAK, MMB, and NVS observed almost the same pulse as GUA, but with higher amplitudes. The stations NUR and BOX located in the northern-dawn region observed pulses that peaked at 0456 and 0457 UT, respectively. In the same longitudinal range, a wide hump of d$H$ with a maximum from 0455 to 0457 UT was detected by the mid-latitude station UPS and the low-latitude station QSB. Note that the latter observed neither a leading nor a trailing decrease in d$H$. The latest observation of a prominent DPD pulse, peaking at 0457 UT, was found at the mid-latitude stations SHU, VIC, and NEW located outside of the transition corridor in the northern-dusk region.

Figure 6b shows the magnetic variations observed during the dawn sector (CZT, HER, HBK, TSU, BNG, TAM, MBO, GUI, HAD, and VAL), and during the dayside region below the lower boundary of the transition corridor (TAN, PAF, AMS, LRM, and GNA). We did not determine a prominent pulse around 0455 UT at all of these stations, with the exception of LRM, which detected an increase of d$H$ from 0458 to 0459 UT. The origin of this hump is unclear. Note that another station, GNA, located very close to LRM, did not detect a pulse. The mid-latitude stations VAL, HAD, and CZT also displayed irregular magnetic variations with amplitudes of a few nT that were much weaker than the pulse of d$H$ observed at the mid-latitude stations UPS, NUR, BOX, and others (see Figure 6a).

INTERMAGNET magnetic observations, with the help of THEMIS, allowed us to determine the following issues:

1. The DPD pulse observed by the radially conjugated ground-based magnetometer were very similar to magnetic variations related to the ECE sequence observed by THEMIS.
2. The magnetic pulse propagated from the magnetopause to the ground during ~1 min or even less.
3. At the ground, the magnetic pulse was first seen in the southern-dusk region then 2 minutes later in the subsolar regions and at low latitudes, and between 3 and 5 minutes in most portions of the mid-latitude Northern Hemisphere. The magnetic pulse did not appear in the dawn sector and in the prenoon and noon regions of the Southern Hemisphere.

While issues 1 and 2 do not contradict existing models [*Chi et al.*, 2006; *Samsonov et al.*, 2011], issue 3 is difficult to explain in the context of MHD waves propagating from a point source through the magnetosphere. The observed pattern is also different from the geomagnetic variation generated by a tilted interplanetary shock gradually covering the dayside magnetopause [*Takeuchi et al.*, 2002]. Therefore, it is reasonable to assume that the DPD pulse somehow relates to the interplanetary discontinuity.

## 5. Discussion and Summary

Using THEMIS observations, in this work, we determined a complex ECE magnetopause distortion. The compression is caused by a very dense, cold, and fast plasma flow with a low magnetic field, called the magnetosheath plasma jet. ECE-related distortions of the magnetopause surface current generate the DPD geomagnetic variations observed by THEMIS in the magnetosphere and by ground-based magnetometers. Using magnetic data from the INTERMAGNET network, we found that the pulse was propagating from the dusk sector of the Southern Hemisphere to the Northern Hemisphere.

Such a pattern of magnetopause disturbances is different from the prediction for FTE, which propagates to both hemispheres from a reconnection line lying near the equatorial plane [*Elphic and Southwood*, 1987]. Prior to the IMF rotation at ~0447 UT (see Figure 3), the IMF orientation yielded a reconnection tailward of the cusps. Afterward, the IMF orientation was compatible with a low-latitude reconnection on the dawn side of the magnetopause. THEMIS, located in the postnoon sector, was remote from the reconnection site.

On the other hand, the $B_n$ component of THEMIS demonstrated bipolar +/- variations at ~0447 and 0455 UT (see lower panels of Figure 2). The magnetic pattern may resemble a magnetospheric FTE, whose appearance is quite possible for a southward IMF [e.g. *Elphic*, 1995]. However, the THEMIS observation of the +/- $B_n$ signature is not consistent with the -/+ FTE signature in the Southern Hemisphere. In addition, the magnetopause exhibited features of a tangential discontinuity that was not appropriate for the reconnection.

In the present case, variations of $B_n$ may result from magnetopause distortions of the type "bulge – indentation – bulge". Namely, the +/- $B_n$ signature at 0447 UT corresponded to an expanding magnetopause with the normal tilted essentially toward north and dusk. The inclination of the magnetopause surface current relative to the nominal orientation caused a local deformation of the geomagnetic field such that the $B_n$ component had a negative variation. At 0454 UT, the magnetopause was locally compressed and undergowent an indentation with the normal tilted southward, as observed by THEMIS during inbound magnetopause crossings between 0454 and 0455 UT. For this case, the inclined magnetopause current caused a strong positive $B_n$ variation. At the trailing edge of the indentation, the inclination of the magnetopause normal changed to northward and resulted in a fast and strong negative variation of the $B_n$ component, as observed by THB from 0455 to 0457 UT.

Therefore, it is unlikely that the observed magnetopause distortions originated from the FTE. Also doubtful is whether the observed high-$\beta$ plasma jet and magnetopause undulation resulted due to the low-$\beta$ magnetic flux ropes (magnetic islands) associated with Kelvin – Helmholtz waves [*Eriksson et al.*, 2009].

The spatial pattern of geomagnetic variations resembled the magnetic effect of pressure pulses generated in the interaction of interplanetary discontinuities and foreshock structures within the magnetosphere [e.g. *Cable and Lin*, 1998; *Sibeck et al.*, 2003; *Turner et al.*, 2011]. Namely, the strength of the pressure variations was weak during the beginning of the interaction, with a tilted interplanetary front. The strongest effects occurred in the hemisphere that was opposite the arrival region.

Structures such as HFA and foreshock cavities are characterized by the magnetic signature of a type of "compression – expansion – compression" [*Sibeck et al.*, 1999; 2002]. Recent reports of near-flank magnetopause distortions indicate that the HFA or a foreshock cavity can only result in a part of a "compression – expansion" type of signature [*Jacobson et al.*, 2009; *Turner et al.*, 2011]. All of these types of structures are different from the ECE sequence. In addition, the necessary distinctive feature of these structures is a rarefaction that results in the prominent expansion of the magnetopause. For our case, INTERMAGNET observations (see Figure 5) indicated that the magnetic pulse is not always accompanied by a decrease in the magnetic field (i.e. by the magnetopause expansion).

The characteristics of the magnetosheath plasma jet observed by THEMIS (see Figures 1 and 2) resembled a TDE generated by interactions of an RD with the bow shock [*Cable and Lin*, 1998; *Tsubouchi and Matsumoto*, 2005]. TDEs displayed an increase for plasma density and a substantial magnetic field decrease in the magnetosheath. However, for our case, the ~ 30 sec temporal scale of the TDE reported by *Hubert and Harvey* [2000] was much smaller than the duration of the plasma jet. In addition, the compressed region of TDEs was not flanked by rarefactions.

We did not find any existing mechanisms that explained the ECE sequence, and it is unclear how the IMF rotation, within ~ 1 min, generated the ECE sequence that lasted for ~ 15 min and that involved a large portion of the dayside magnetosphere. Below, we discuss possible scenarios that may explain THEMIS observations for the ECE sequence and INTERMAGNET observations of DPD pulses.

For our case, DD slanted along the dayside bow shock from southern-dawn to northern-dusk. Near the magnetopause, THEMIS observed very strong pressure pulses, or jets, in the postnoon sector. At the THEMIS location, the magnetopause distance was 12.2 Re and the distance to the bow shock was 17.3 Re. By taking a $V_n$ = -100 km, as observed by the THE in the jet, we estimated that the jet propagated from the bow shock to the magnetopause in approximately ~5 min. However, we did not have experimental information regarding jet generation at the bow shock. The actual travel time could be longer, in cases of non-radial propagation, or shorter, due to the deceleration of the jet from higher speeds at the bow shock to $V_n$~100 km/s near the magnetopause. Note that a similar travel time of ~5 to 6 min was obtained by *Eastwood et al.* [2011] for the radial propagation of a magnetosheath perturbation generated during the interaction of the HFA with the flank bow shock, and by *Villante et al.* [2004] for a statistical analysis of geomagnetic disturbances produced by interplanetary shocks.

For the first approach, we assume that the jet, observed by THEMIS at 0455 UT, was launched from the bow shock at ~0450 UT. In Figure 5, the time corresponds to a DD propagation time of d$T$ ~ 49 min, whose isocontour passed very close to the THEMIS location. Within a few-minute uncertainty of the DD propagation time, we suggest that the jet is related to the interplanetary discontinuity.

Regarding the DD propagation time d$T$, we found that at early times of DD arrival to the bow shock in the southern-dawn sector, no magnetic pulse was detected. Hence, the magnetosheath pressure pulse was not generated when the DD was approaching and began to interact with the bow shock. The generation of the d$H$ pulse began ~10 – 12 minutes after DD arrival. On the other hand, the simultaneous (within 1 min) occurrence of the pulse at stations PHU and MMB indicated that magnetopause compression did not directly follow the trace of the DD front along the bow shock. Otherwise, the magnetic pulses detected by those stations should be separated by ~5 min, which is not the case.

Simultaneous observations of the pulse at widely separated stations can be explained by the generation of the jet at the subsolar bow shock following propagation along magnetosheath streamlines. However, the latter scenario seems unlikely since for this case, the jet should be observed in any direction with a time delay that increases from the subsolar region. Figure 5 absolutely demonstrates a different pattern. A magnetic pulse was first seen in the southern-dusk region, then in the subsolar region and at low GSM latitudes on the dayside, and later in the middle latitudes of the Northern Hemisphere.

As shown in Figure 4, the magnetic pulse timing partially resembled the pattern of a $Q_\parallel$ bow shock transition from the south-dusk to the north-dawn sector. The transition began in the southern-dusk region, where the original $Q_\parallel$ portion of the bow shock was gradually replaced in ~10 min by $Q_\perp$. In this region and at the same time, the magnetic stations EYR, CNB, and ASP detected the earliest enhancement of d$H$ that lasted up to 10 min (see Figure 6a). After ~10 min, the transition finished in the northern-dawn sector, where the $Q_\perp$ portion of the bow shock was replaced by $Q_\parallel$. In this region, the DPD pulse was detected by stations NUR and UPS for the latest time of ~0457 UT.

Note that the DPD pulse was also observed at the later time of ~0457 UT by stations SHU, VIC, and NEW located in the dusk sector at northern geomagnetic latitudes of ~55°. Therefore, it is difficult to explain the time of magnetic pulse appearance at these stations in the context of the DD propagation time (d$T$ ~ 60 min in contrast to ~49 min at THEMIS), as well as by the propagation of the pressure pulse from the subsolar region along magnetosheath streamlines that take at least 6 min with a speed of ~300 km/s. We assume that these mid-latitude stations observed the magnetic effect of the ionospheric currents induced at higher latitudes by Alfvén waves arriving from magnetopause distortions at low-latitudes.

Generation of the pulse in the south-dusk region can be explained by the model of TDE [*Lin et al.*, 1996]. The model predicts the generation of a magnetosheath pressure pulse through interactions of the bow shock with the RD. The pulse is very weak in the beginning of interaction and becomes stronger with RD propagation along the bow shock. The pulse is most prominent within a region where the bow shock is translated from $Q_\parallel$ to $Q_\perp$ (i.e. in the region of a collapsing foreshock boundary). In this region, an upstream pressure pulse is generated by suprathermal ions reflected in the foreshock region, when the IMF is rotated from a $Q_\parallel$ to a $Q_\perp$ orientation. The upstream pulse propagates quickly toward the magnetopause, catches up, and enforces the magnetosheath pressure pulse. *Lin et al.* [1996] found that the foreshock pulse may have a spatial scale of several earth radii, corresponding to the pulse duration of a few minutes. Hence, a portion of RD interacting with the bow shock in the region of a collapsing foreshock should be accompanied by a strong and large-scale density enhancement in the magnetosheath plasma.

For our case, DD was quite similar to RD. The collapsing foreshock was located in the south-dusk region where the motion of DD led to the gradual transition of the bow shock from $Q_\parallel$ to $Q_\perp$. The transition should result in the generation of a magnetosheath pressure pulse and, as a consequence, a compression of the magnetopause. In the same region, the ground-based stations EYR, CNB, ASP, CTA, and KDU observed a prominent magnetic hump beginning at ~0451 UT. As a result of the ~5 min travel time for the pressure pulse in the magnetosheath, we estimated that the pulse began from the bow shock at ~0446 UT. The time corresponds to the iso-contour of the DD propagation time, d$T$ = 45 min. Within a-few-minutes in accuracy, this iso-contour was very close to the location of the stations (see Figure 5). Additionally, as shown in Figures 4 and 5, the duration of the magnetic hump of up to ~10 min mostly corresponded to the time of DD sliding across the $Q_\parallel$ portion of the bow shock (from d$T$ ~ 44 min to d$T$ ~ 54 - 60 min). Note that after d$T$ ~ 54 min (0455 UT) only a minor portion of the $Q_\parallel$ bow shock was left in the south-dusk region. Therefore, the foreshock pulse began to diminish. So, we suggest that stations located in the south-dusk region observed the net magnetic effect of the magnetopause compression caused by the large-scale magnetosheath pressure pulse that is generated in the interaction of DD, with the bow shock accompanied by foreshock collapsing.

At low latitudes on the dayside, the model of TDE predicts the gradual propagation (during ~5 min) of the magnetosheath pressure pulse across the transition corridor from south-dawn to northern dusk. The pulse should be oriented along the iso-contours of d$T$ (i.e. along the line of the DD interaction with the bow shock (see Figure 5)). We found that the THEMIS and INTERMAGNET stations located around the subsolar point almost simultaneously observed the magnetic pulse that peaked at ~0455 UT. The model of TDE seems to fails in this region.

From Figure 4 one can see that the region of ~40° surrounding the subsolar point was characterized by the fast translation of the $Q_\parallel$ portion of the bow shock, from south-dusk to north-dawn, when the DD was sliding across the transition corridor. The translation occurred along the iso-contours of d$T$ (see Figure 5), corresponding to the line of the DD interaction with the bow shock. Therefore, we believe that the jet is somehow generated during this translation. The THEMIS location is very close to the iso-contour of d$T$ = 50 min that passes the center of the $Q_\parallel$ portion of the bow shock in the south-dusk sector. Therefore, it is quite possible that the foreshock collapsing in that region (where the IMF is almost parallel to the bow shock) results in the most intense magnetosheath pressure pulse. The pulse generation region may follow the portion of the parallel bow shock, translating quickly to the northern dawn sector and passing THEMIS on the way. The jet observed by THEMIS could be explained if we assumed that the strong pulse was generated continuously during the translation. The assumption also allowed us to explain the northward direction of the jet propagation observed by THEMIS.

Using this assumption, the duration of the jet was determined by the time of the IMF rotation. In Figure 3, we found that at ACE, the IMF rotated during ~40 sec from 04:00:10 to 04:00:50 UT. THEMIS observed a jet with a duration of ~70 sec. Therefore, the DD could be widening during its propagation from ACE to the Earth, or during interactions with the foreshock region.

Another problem is the simultaneous appearance of a magnetic pulse that peaks at ~0455 UT in the range of GSM longitudes (from ~-20° to ~60°) and latitudes (from -10° to 40°). Here, we must take into account the fact that the ground-based stations observed a net magnetic effect. The most prominent magnetic enhancement was produced by a jet with a ~70 sec duration. Because the jet is translated quickly (~3 min), it is quite possible that within a 1 min time resolution, the intense peak of jet-associated magnetic enhancements can be observed at almost the same time by widely separated stations. Taking a jet translation time of ~3 min and an angular size of the translation of ~120°, we determined that the jet has a very large spatial scale of ~10 Re.

Note that the above scenario does not exclude the generation of the magnetosheath pressure pulse in the DD interaction with the bow shock as predicted by the model of TDE. However, the magnetic effect from this pulse may be masked by the strong magnetic enhancement produced by the jet. In Figure 6a we determined that in the subsolar region stations PHU, GZH, and LZH did not observe any pulses before 0453 UT, corresponding to the iso-contour of $dT = 47$ min. Here, we again assumed a 5 min travel time in the magnetosheath. The iso-contour was located very close to the stations and it seems that the DD-related magnetosheath pulse (if any) was still very weak. The result is indirect support of our hypothesis that low-latitude stations observe the magnetic signal from a remote source. In the same manner, the signal could possibly be observed at stations KNY, KAK, MMB, and NVS.

From Figure 4c one can see that the translating portion of the bow $Q_\parallel$ shock was replaced by $Q_\perp$ only until the region where the angle $\theta_{Bn}$ became smaller than 45°. The region is located not so far from the subsolar point in the south-dawn direction. Therefore, it is reasonable to assume that in this region the generation of the pressure pulse diminished and that the mid-latitude stations NUR and UPS observed a relatively weak pulse that peaked at ~0457 UT. The HAD, VAL, GUI, MBO, and TAM stations did not detect a practical pulse.

The magnetic signal from the leading magnetopause expansion of the ECE sequence observed by THEMIS was revealed as a prominent magnetic depletion preceding the pulse at the stations located above the iso-contour of $dT = 48$ min (see Figure 5) (i.e. after DD crossed the subsolar bow shock). The decrease may result from a magnetopause expansion caused by the radial IMF imposing a subsolar bow shock [e.g. *Suvorova et al.*, 2010]. However, the magnetopause expansion should be observed by surrounding stations such as LZH, PHU, and GZH which was not the case. In addition, at the THEMIS location, the bulge of the expanded magnetopause had an orientation from south-dawn to north-dusk that is not consistent with the propagation of the expansion from the subsolar region. Hence, the leading expansion may be part of a complex magnetopause distortion produced by a quickly translating large-scale jet. The hypothesis is supported by observations of a preceding depletion at stations located northward and dawnward from THEMIS (i.e. in the way of the jet translation).

The trailing expansion of the ECE sequence, observed at many magnetic stations as a trailing decrease in the DPD pulse, may result from two different effects. The non-stationary process of jet interaction with the magnetopause may lead to the development of a plasma turbulence in the trailing region that results in magnetopause undulations, as observed by the THA probe. Another possible effect is magnetopause expansion due to a reflected fast shock or a fast wave [*Samsonov et al.*, 2007]. The wave propagates sunward and follows magnetopause compression within a few minutes that corresponded to the observed dynamics of the magnetopause expansion. Further experimental and modeling studies of these phenomena are required in order to understand the nature of the ECE sequence.

With the help of THEMIS observations, we also found very important evidence that jet interactions with the magnetopause can be accompanied by the effective penetration of magnetosheath plasma inside the magnetosphere. Most of the intense magnetosheath plasma streams were revealed in the magnetosphere during the post-jet expansion of the magnetopause. The thickness of the magnetopause was estimated to be close to a nominal ~ 500 km. Hence, the kinetic mechanism of ion penetration through the thin (~ 90 km) current sheet does not work for the present case. Here, we speculate that the penetration mechanism may be similar to that operating under a KH instability, when the magnetopause shape is dramatically distorted. Further studies using sophisticated modeling simulations are required.

In summary, we suggest that the large-scale (~10 Re) magnetosheath plasma jet, observed in-situ by THEMIS, can be generated by interactions of the interplanetary directional discontinuity with the bow shock. The plasma jet produces a prominent magnetopause distortion that results in the effective penetration of the magnetosheath plasma inside the magnetosphere. The jet-associated magnetopause compression is preceded and followed by magnetopause expansions (ECE sequence). Strong magnetopause undulations of the ECE type with the amplitude of ~1.5 Re generate a tripolar magnetic pulse "decrease – peak – decrease", observed at low and middle latitudes by an INTERMAGNET network of ground-based magnetometers. Pulse observations allowed us to trace the magnetopause distortion traveling within the bow shock translation region, that is guided by the directional discontinuity sliding along the dayside magnetosphere. The mechanism of jet generation and the formation of the ECE sequence was found to be unclear. We speculate that the jet may be generated in a localized region of bow shock transition from the $Q_\parallel$ to $Q_\perp$ orientation, which translates quickly from one equilibrium location to another.

**Acknowledgements** We acknowledge NASA contract NAS5-02099 and V. Angelopoulos for use of data from the THEMIS mission. We thank K.H. Glassmeier and U. Auster for the use of FGM data provided under contract 50 OC 0302 and C. W. Carlson and J. P. McFadden for use of ESA data. We thank N. Ness and D.J. McComas for the use of ACE solar wind data made available via the CDAWeb. The results presented in this paper rely on data collected at magnetic observatories. We thank the national institutes that support them and INTERMAGNET for promoting high standards of magnetic observatory practice (www.intermagnet.org). We also thank Kyoto World Data Center for Geomagnetism (http://swdcwww.kugi.kyoto-u.ac.jp/index.html) for providing the geomagnetic indices. We appreciate J.-K. Chao for the contribution of useful discussion and comments. We are grateful to the reviewers for their very fruitful comments. This work was supported by grants NSC-99-2111-M-008-013 and NSC 99-2811-M-008-093 from the National Science Council of Taiwan and by Ministry of Education under the Aim for Top University program at National Central University of Taiwan #985603-20.

# References

Amata, E., et al. (2006), Experimental study of nonlinear interaction of plasma flow with charged thin current sheets: 1. Boundary structure and motion, *Nonlin. Processes Geophys.*, *13*, 365–376.

Amata, E., S. P. Savin, D. Ambrosino, Y. V. Bogdanova, M. F. Marcucci, S. Romanov, A. Skalsky (2011), High kinetic energy density jets in the Earth's magnetosheath: A case study, *Planet. Space Sci.*, *59*, 482–494, doi:10.1016/j.pss.2010.07.021.

Auster, H. U., et al. (2008), The THEMIS fluxgate magnetometer, *Space Sci. Rev.*, *141*, 235–264, doi:10.1007/s11214-008-9365-9.

Berchem, J., and C. Russell (1982), The thickness of the magnetopause current layer: ISEE 1 and 2 observations, *J. Geophys. Res.*, *87*(A4), 2108–2114.

Berchem, J., and C. Russell (1984), Flux transfer events on the magnetopause: Spatial distribution and controlling factors, *J. Geophys. Res.*, *89*(A8), 6689–6703.

Cable, S., and Y. Lin (1998), Three-dimensional MHD simulations of interplanetary rotational discontinuities impacting the Earth's bow shock and magnetosheath, *J. Geophys. Res.*, *103*(A12), 29,551–29,567.

Chao, J. K., D. J. Wu, C.-H. Lin, Y. H. Yang, and X. Y. Wang (2002), Models for the size and shape of the Earth's magnetopause and bow shock, *in Space Weather Study Using Multipoint Techniques*, edited by L.-H. Lyu, 360 pp., Pergamon, New York, 2002.

Chi, P. J., D.-H. Lee, and C. T. Russell (2006), Tamao travel time of sudden impulses and its relationship to ionospheric convection vortices, *J. Geophys. Res.*, *111*, A08205, doi:10.1029/2005JA011578.

Dmitriev, A.V., and H.-C. Yeh (2008), Geomagnetic signatures of sudden ionospheric disturbances during extreme solar radiation events, Spec. Issue "Ionospheric effects and telecommunications", Eds. A. Bourdillon, B. Zolesi, A. Dmitriev, *J. Atmos. Sol.-Terr. Phys.*, *70*, 1971–1984, doi:10.1016/j.jastp.2008.05.008.

Eastwood, J. P., E. A. Lucek, C. Mazelle, K. Meziane, Y. Narita, J. Pickett, and R. Treumann (2005), The foreshock, *Space Science Reviews*, *118*(1-4), 41–94, doi:10.1007/s11214-005-3824-3.

Eastwood, J. P., et al., (2011), Transient Pc3 wave activity generated by a hot flow anomaly: Cluster, Rosetta, and ground-based observations, *J. Geophys. Res.*, 116, A08224, doi:10.1029/2011JA016467.

Elphic, R.C. (1995), Observations of flux transfer events: A review, in *Physics of the Magnetopause, AGU Monograph* 90, 225–233.

Elphic, R., and D. Southwood (1987), Simultaneous measurements of the magnetopause and flux transfer events at widely separated sites by AMPTE UKS and ISEE 1 and 2, *J. Geophys. Res.*, *92*(A12), 13,666–13,672.

Eriksson, S., et al. (2009), Magnetic island formation between large-scale flow vortices at an undulating postnoon magnetopause for northward interplanetary magnetic field, *J. Geophys. Res.*, *114*, A00C17, doi:10.1029/2008JA013505.

Facskó, G., Z. Németh, G. Erdös, A. Kis, and I. Dandouras (2009), A global study of hot flow anomalies using Cluster multi-spacecraft measurements, *Ann. Geophys.*, *27*, 2057–2076.

Fairfield, D., A. Otto, T. Mukai, S. Kokubun, R. Lepping, J. Steinberg, A. Lazarus, and T. Yamamoto (2000), Geotail observations of the Kelvin-Helmholtz instability at the equatorial magnetotail boundary for parallel northward fields, *J. Geophys. Res.*, *105*(A9), 21,159–21,173.

Friis-Christensen, E., M. A. McHenry, C. R. Clauer, and S. Vennerstrom (1988), Ionospheric traveling convection vortices observed near the polar cleft: A triggered response to sudden changes in the solar wind, *Geophys. Res. Lett.*, *15*, 253–256.

Hasegawa, H., B. U. Ö. Sonnerup, C. J. Owen, B. Klecker, G. Paschmann, A. Balogh, and H. Rème (2006), The structure of flux transfer events recovered from Cluster data, *Ann. Geophys.*, *24*, 603–618.

Hubert, D., and C. C. Harvey (2000), Interplanetary rotational discontinuities: From the solar wind to the magnetosphere through the magnetosheath, *Geophys. Res. Lett.*, *27*(19), 3149–3152.

Jacobsen, K. S., et al. (2009), THEMIS observations of extreme magnetopause motion caused by a hot flow anomaly, *J. Geophys. Res.*, *114*, A08210, doi:10.1029/2008JA013873.

Kawano, H., S. Kokubun, and K. Takahashi (1992), Survey of transient magnetic field events in the dayside magnetosphere, *J. Geophys. Res.*, *97*(A7), 10,677–10,692.

Lin, Y. (2002), Global hybrid simulation of hot flow anomalies near the bow shock and in the magnetosheath, *Planet. Space Sci.*, *50*, 577–591.

Lin, Y., D. Swift, and L. Lee (1996), Simulation of pressure pulses in the bow shock and magnetosheath driven by variations in interplanetary magnetic field direction, *J. Geophys. Res.*, *101*(A12), 27,251–27,269.

Lin, Y., and X. Wang (2005), Three-dimensional global hybrid simulation of dayside dynamics associated with the quasiparallel bow shock, *J. Geophys. Res.*, *110*, A12216, doi:10.1029/2005JA011243.

Lin, C. C., C. L. Tsai, H. J. Chen, C. J. Weng, J. K. Chao, and L. C. Lee (2009), A possible generation mechanism of interplanetary rotational discontinuities, *J. Geophys. Res.*, *114*, A08102, doi:10.1029/2008JA014008.

Lin, R. L., X. X. Zhang, S. Q. Liu, Y. L. Wang, and J. C. Gong (2010), A three-dimensional asymmetric magnetopause model, *J. Geophys. Res.*, *115*, A04207, doi:10.1029/2009JA014235.

McFadden, J. P., et al. (2008), The THEMIS ESA plasma instrument and in-flight calibration, *Space Sci. Rev.*, *141*, 277–302, doi:10.1007/s11214-008-9440-2.

Němeček, Z., et al. (1998), Transient flux enhancements in the magnetosheath, *Geophys. Res. Lett.*, *25*(8), 1273–1276.

Neugebauer, M., D. Clay, B. Goldstein, B. Tsurutani, and R. Zwickl (1984), A reexamination of rotational and tangential discontinuities in the solar wind, *J. Geophys. Res.*, *89*(A7), 5395–5408.

Omidi, N., and D. G. Sibeck (2007), Formation of hot flow anomalies and solitary shocks, *J. Geophys. Res.*, *112*, A01203, doi:10.1029/2006JA011663.

Russell, C. T., and S. M. Petrinec (1996), Comment on ''Towards an MHD theory for the standoff distance of Earth's bow shock'' by I.H. Cairns and C.L. Grabbe, *Geophys. Res. Lett.*, 23, 309–310.

Samsonov, A. A., D. G. Sibeck, and J. Imber (2007), MHD simulation for the interaction of an interplanetary shock with the Earth's magnetosphere, *J. Geophys. Res.*, *112*, A12220, doi:10.1029/2007JA012627.

Samsonov, A. A., et al. (2011), Propagation of a sudden impulse through the magnetosphere initiating magnetospheric Pc5 pulsations, *J. Geophys. Res.*, 116, A10216, doi:10.1029/2011JA016706.

Savin, S., et al. (2004), Dynamic interaction of plasma flow with the hot boundary layer of a geomagnetic trap, *JETP Lett.*, *79*(8), 368–371.

Savin, S., et al. (2006), Experimental study of nonlinear interaction of plasma flow with charged thin current sheets: 2. Hall dynamics, mass and momentum transfer, *Nonlin. Processes Geophys.*, *13*, 377–392.

Savin, S., et al. (2008), High kinetic energy jets in the Earth's magnetosheath: Implications for plasma dynamics and anomalous transport, *JETP Lett.*, *87*(11), 593–599.

Sibeck, D. G., N. L. Borodkova, G. N. Zastenker, S. A. Romanov, and J. -A. Sauvaud (1998), Gross deformation of the dayside magnetopause, *Geophys. Res. Lett.*, *25*(4), 453–456.

Sibeck, D. G., et al. (1999), Comprehensive study of the magnetospheric response to a hot flow anomaly, *J. Geophys. Res.*, *104*(A3), 4577–4593.

Sibeck, D. G., et al. (2000), Magnetopause motion driven by interplanetary magnetic field variations, *J. Geophys. Res.*, *105*(A11), 25,155–25,169.

Sibeck, D. G., T.-D. Phan, R. Lin, R. P. Lepping, and A. Szabo (2002), Wind observations of foreshock cavities: A case study, *J. Geophys. Res.*, *107*(A10), 1271, doi:10.1029/2001JA007539.

Sibeck, D. G., et al. (2003), Pressure-pulse interaction with the magnetosphere and ionosphere, *J. Geophys. Res.*, *108*(A2), 1095, doi:10.1029/2002JA009675.

Sibeck, D. G., et al. (2008), Crater FTEs: Simulation results and THEMIS observations, *Geophys. Res. Lett.*, *35*, L17S06, doi:10.1029/2008GL033568.

Sonnerup, B., I. Papamastorakis, G. Paschmann, and H. Luhr (1987), Magnetopause properties from AMPTE/IRM observations of the convection electric field: Method development, *J. Geophys. Res.*, *92*(A11), 12,137–12,159.

Suvorova, A. V., et al. (2010), Magnetopause expansions for quasi-radial interplanetary magnetic field: THEMIS and Geotail observations, *J. Geophys. Res.*, *115*, A10216, doi:10.1029/2010JA015404.

Takeuchi, T., C. T. Russell, and T. Araki (2002), Effect of the orientation of interplanetary shock on the geomagnetic sudden commencement, *J. Geophys. Res.*, *107*(A12), 1423, doi:10.1029/2002JA009597.

Teh, W.-L., B. U. Ö. Sonnerup, G. Paschmann, and S. E. Haaland (2011), Local structure of directional discontinuities in the solar wind, *J. Geophys. Res.*, *116*, A04105, doi:10.1029/2010JA016152.

Thomas, V., and S. Brecht (1988), Evolution of Diamagnetic Cavities in the Solar Wind, *J. Geophys. Res.*, *93*(A10), 11,341–11,353.

Thomsen, M., J. Gosling, S. Fuselier, S. Bame, and C. Russell (1986), Hot, diamagnetic cavities upstream from the Earth's bow shock, *J. Geophys. Res.*, *91*(A3), 2961–2973.

Tsubouchi, K., and H. Matsumoto (2005), Effect of upstream rotational field on the formation of magnetic depressions in a quasi-perpendicular shock downstream, *J. Geophys. Res.*, *110*, A04101, doi:10.1029/2004JA010818.

Turner, D. L., et al. (2011), Multispacecraft observations of a foreshock-induced magnetopause disturbance exhibiting distinct plasma flows and an intense density compression, *J. Geophys. Res.*, *116*, A04230, doi:10.1029/2010JA015668.

Villante, U., S. Lepidi, P. Franciaa, T. Brunoa (2004), Some aspects of the interaction of interplanetary shocks with the Earth's magnetosphere: an estimate of the propagation time through the magnetosheath, *Journal of Atmospheric and Solar-Terrestrial Physics*, 66, 337-341.

Weimer, D. R., et al. (2002), Variable time delays in the propagation of the interplanetary magnetic field, *J. Geophys. Res.*, *107*(A8), doi:10.1029/2001JA009102.

**Table 1.** GSM location of the THEMIS probes on June 16, 2007

| Probe | 0440 UT | | 0510 UT | |
|---|---|---|---|---|
| | R, Re | X, Y, Z | R, Re | X, Y, Z |
| THA | 12.8 | 8.1, 9.5, -2.8 | 12.5 | 7.7, 9.4, -3.0 |
| THE | 12.0 | 7.0, 9.4, -2.7 | 11.7 | 6.6, 9.2, -2.8 |
| THC | 11.9 | 6.9, 9.3, -2.6 | 11.5 | 6.4, 9.1, -2.7 |
| THD | 11.8 | 6.8, 9.3, -2.6 | 11.4 | 6.4, 9.1, -2.7 |
| THB | 11.3 | 6.2, 9.1, -2.5 | 10.8 | 5.7, 8.8, -2.5 |

**Table 2.** Magnetopause crossings and normal in GSM coordinates on June 16, 2007

| Probe | MVA interval, UT | Type | X, Y, Z, Re | eigenvectors | eigenvalues |
|---|---|---|---|---|---|
| THA | 0446:30 - 0448:00 | outbound | 8.0, 9.5, -2.9 | 0.31, 0.93, 0.19 | 334, 23, 7 |
| THA | 0452:00 - 0453:30 | inbound | 7.9, 9.5, -2.9 | 0.76, -0.40, -0.50 | 612, 56, 6 |
| THE | 0454:00 - 0455:00 | inbound | 6.8, 9.3, -2.7 | 0.63, 0.68, -0.37 | 1450, 31, 2 |
| THC | 0453:30 - 0455:00 | inbound | 6.7, 9.2, -2.7 | 0.11, 0.98, -0.16 | 806, 48, 4 |
| THD | 0453:30 - 0455:30 | inbound | 6.6, 9.2, -2.6 | 0.28, 0.92, -0.26 | 1200, 71, 12 |
| THD | 0456:00 - 0457:00 | outbound | 6.6, 9.1, -2.7 | 0.10, 0.99, 0.10 | 671, 32, 5 |
| THC | 0456:00 - 0457:00 | outbound | 6.6, 9.2, -2.7 | 0.17, 0.98, 0.08 | 790, 30, 5 |
| THE | 0456:00 - 0457:30 | outbound | 6.8, 9.3, -2.7 | 0.07, 0.99, 0.03 | 742, 20, 6 |
| THA | ~0459:00 | outbound | 7.9, 9.4, -3.0 | - | - |
| THA | ~0500:06 | inbound | 7.9, 9.4, -3.0 | - | - |
| THA | ~0500:25 | outbound | 7.8, 9.4, -3.0 | - | - |
| THA | ~0500:57 | inbound | 7.8, 9.4, -3.0 | - | - |

**Table 3.** Location of magnetic stations in geomagnetic coordinates

| Code | Name | mlon | mlat |
|---|---|---|---|
| AAA | Alma Ata | 151.35 | 33.41 |
| AAE | Addis Ababa | 109.87 | 5.34 |
| AMS | Martin de Vivies-Amsterdam Island | 141.30 | -47.09 |
| ASP | Alice Springs | -153.62 | -34.14 |
| BNG | Bangui | 89.28 | 4.53 |
| BOX | Borok | 123.74 | 52.90 |
| CNB | Canberra | -134.44 | -43.75 |
| CTA | Charters Towers | -140.68 | -29.23 |
| CZT | Port Alfred | 109.87 | -51.40 |
| EYR | Eyrewell | -106.99 | -47.77 |
| GNA | Gnangara | -173.24 | -43.13 |
| GUA | Guam | -146.30 | 4.16 |
| GUI | Guimar-Tenerife | 59.23 | 34.65 |
| GZH | Zhaoqing | -177.93 | 11.61 |
| HAD | Hartland | 79.67 | 54.47 |
| HBK | Hartebeesthoek | 91.85 | -26.83 |
| HER | Hermanus | 81.33 | -33.43 |
| IRT | Irkutsk | 175.38 | 40.85 |
| KAK | Kakioka | -153.29 | 26.16 |
| KDU | Kakadu | -156.25 | -23.20 |
| KNY | Kanoya | -161.19 | 20.67 |
| LER | Lerwick | 89.14 | 62.34 |
| LRM | Learmonth | -175.67 | -33.57 |
| LZH | Lanzhou | 174.31 | 24.76 |
| MBO | Mbour | 55.77 | 21.07 |
| MMB | Memambetsu | -150.86 | 34.17 |
| NEW | Newport | -58.92 | 55.19 |
| NUR | Nurmijarvi | 113.07 | 57.75 |
| NVS | Novosibirsk | 158.52 | 44.39 |
| PAF | Port-aux-Francais | 129.11 | -57.46 |
| PHU | Phuthuy | 175.93 | 9.73 |
| QSB | Qsaybeh | 112.18 | 30.21 |
| SHU | Shumagin | -106.62 | 53.66 |
| SIT | Sitka | -83.67 | 60.18 |
| TAM | Tamanrasset | 80.28 | 25.22 |
| TAN | Antananarivo | 113.56 | -23.84 |
| TSU | Tsumeb | 83.53 | -18.30 |
| UPS | Uppsala | 106.35 | 58.53 |
| VAL | Valentia | 74.21 | 56.42 |
| VIC | Victoria | -66.00 | 54.28 |

**Figure Captions**

Figure 1. Magnetopause disturbances at 0440 – 0510 UT on 16 June 2007 (from top to bottom), as follows: the THE and THB ion and electron spectrograms; the GSM components $Vx$ (black), $Vy$ (red), and $Vz$ (blue) of the ion velocity measured by THE and THB; the magnetic field strength, and the GSM $Bz$ and $By$ components, measured by THA (black), THB (red), THC (blue), THD (green), THE (azure), and ACE (orange). The THEMIS magnetic field was divided by 10. The vertical blue and red dashed lines depict, respectively, the magnetopause outbound and inbound crossings owing to a sequence of expansion – compression – expansion of the magnetopause. During the magnetopause compression at ~0454, fast streams of the magnetosheath plasma (see the fifth panel from the top) penetrated the magnetosphere and remained there until ~0508 UT (see the third and fourth panels from the top).

Figure 2. THEMIS observations of a magnetosheath plasma jet and magnetopause distortions in the frame of the undisturbed nominal magnetopause according to *Lin et al.* [2010] at 0445 – 0502 UT on 16 June 2007. The top panel shows THE data for magnetic (red line), thermal (blue line), dynamic (green line), and total (black line) pressure within the magnetosphere and the magnetosheath, as well as the total solar wind pressure observed by ACE (the gray dashed line). Vertical red and blue dashed lines depict, respectively, the THE outbound and inbound magnetopause crossings caused by a magnetosheath plasma jet from 04:44:10 to 04:46:40 UT. The second panel shows the plasma density (solid lines) and the temperature (dashed lines) measured by THE (azure) and THB (red). The third panel shows the orientation of magnetopause normals in planes **l-n** (red segments) and **m-n** (blue segments). The normal distance of THEMIS probes relative to the nominal magnetopause (the gray thick horizontal line at $R_n = 0$) is shown by thin gray lines. A sketch of magnetopause location is shown by the dashed line. Note that the THB does not enter into the magnetosheath. The fourth through eighth panels show THEMIS observations of the magnetic field, as well as the plasma velocity in nominal magnetopause coordinates. The total values and the components **l**, **m,** and **n** are shown by black, blue, and green and red curves, respectively. The jet causes a substantial distortion in the shape of the magnetopause shape, perturbations in the magnetic field, and plasma fluxes.

Figure 3. The upstream solar wind conditions measured by ACE on June 16, 2007 (from top to bottom), as follows: the IMF components $Bz$ (black curve, left axis) and $By$ (gray curve, right axis) in GSM; the IMF magnitude $B$ (black curve, left axis) and the $Bx$ component (gray curve, right axis); the helium abundance (black curve, left axis) and the proton temperature (gray curve, right axis); the plasma velocity $Vx$-component (solid line, left axis), the $Vy$ and $Vz$ components in GSM (black and gray dotted lines, respectively, right axis); and the sum of the solar wind dynamic, thermal and magnetic pressures (black curve, left axis), and plasma density (gray curve, right axis). The vertical dashed lines restrict the intervals used for the analysis of a directional discontinuity (DD) centered at ~04:00:46 UT.

Figure 4. GSM maps of the contours of equal cone angles $\theta_{Bn}$ (in degrees) between the IMF vector and normal to *Chao et al.'s* [2002] bow shock, calculated for different orientations of the IMF observed by ACE, as follows: (a) at 0355 UT before DD arrival; (b) at ~04:00:46 UT in the center of DD; (c) at 0406 UT, after DD passing. Thick curves restrict the regions of the quasi-parallel bow shock ($\theta_{Bn} < 25°$). The THEMIS location and the subsolar point are indicated by the black cross and the gray circle, respectively. The thick black dotted lines indicate a transitions corridor where the quasi-parallel bow shock and the foreshock transit from the southern-dusk to the northern-dawn region.

Figure 5. A GSM map of the INTERMAGNET stations (triangles) at 0455 UT on 16 June 2007. Four-digit numbers indicate the time moment (UT) of the magnetic peak observation. The letter "d" before (after) the numbers indicates that the peak is preceded (followed) by a decrease in the magnetic field. The absence of geomagnetic signatures is denoted by "####". Gray curves with numbers depict the iso-contours of the DD propagation time from the ACE upstream monitor to *Chao et al.'s* [2002] bow shock. The THEMIS location, the sub-solar point, and the transition corridor are indicated by black crosses, gray circles, and dotted lines, respectively. No prominent magnetic variations occurred during the dawn sector. The magnetic pulse appears in the southern-dusk sector and propagates northward and dawnward.

Figure 6. Variations in the horizontal component (d$H$) of the geomagnetic field, as follows: (a) prominent variations in the dayside and dusk sectors, with the THB magnetic field shown for reference with the black dashed curve, and (b) relatively weak variations in the dawn sector and in the beginning of the DD interaction with a bow shock. The geographic longitude and latitude of the stations are indicated with parentheses. Vertical dashed lines depict the time, 0455 UT, of the magnetic pulse maximum of THEMIS.

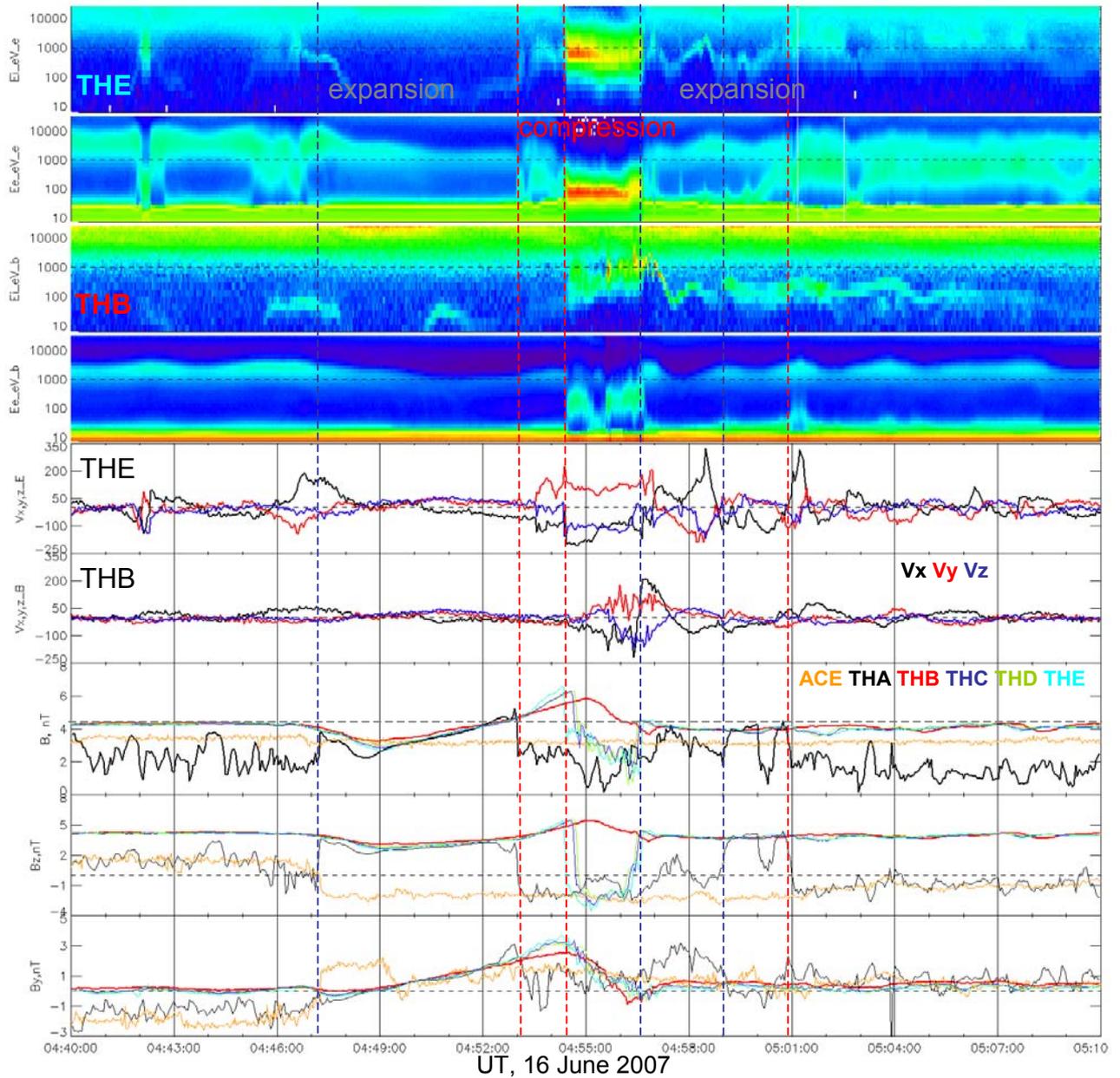

**Figure 1.** Magnetopause disturbances at 0440 – 0510 UT on June 16 2007 (from top to bottom): THE and THB ion and electron spectrograms; GSM components $V_x$ (black), $V_y$ (red) and $V_z$ (blue) of ion velocity measured by THE and THB; magnetic field strength, GSM $B_z$ and $B_y$ components, measured by THA (black), THB (red), THC (blue), THD (green), THE (azure) and ACE (orange). The THEMIS magnetic field is divided by 10. Vertical blue and red dashed lines depict, respectively, magnetopause outbound and inbound crossings owing to a sequence of expansion – compression – expansion of the magnetopause. During the magnetopause compression at ~0454, fast streams of the magnetosheath plasma (see the fifth panel from the top) penetrate inside the magnetosphere and remain there until ~0508 UT (see the third and fourth panels from the top)..

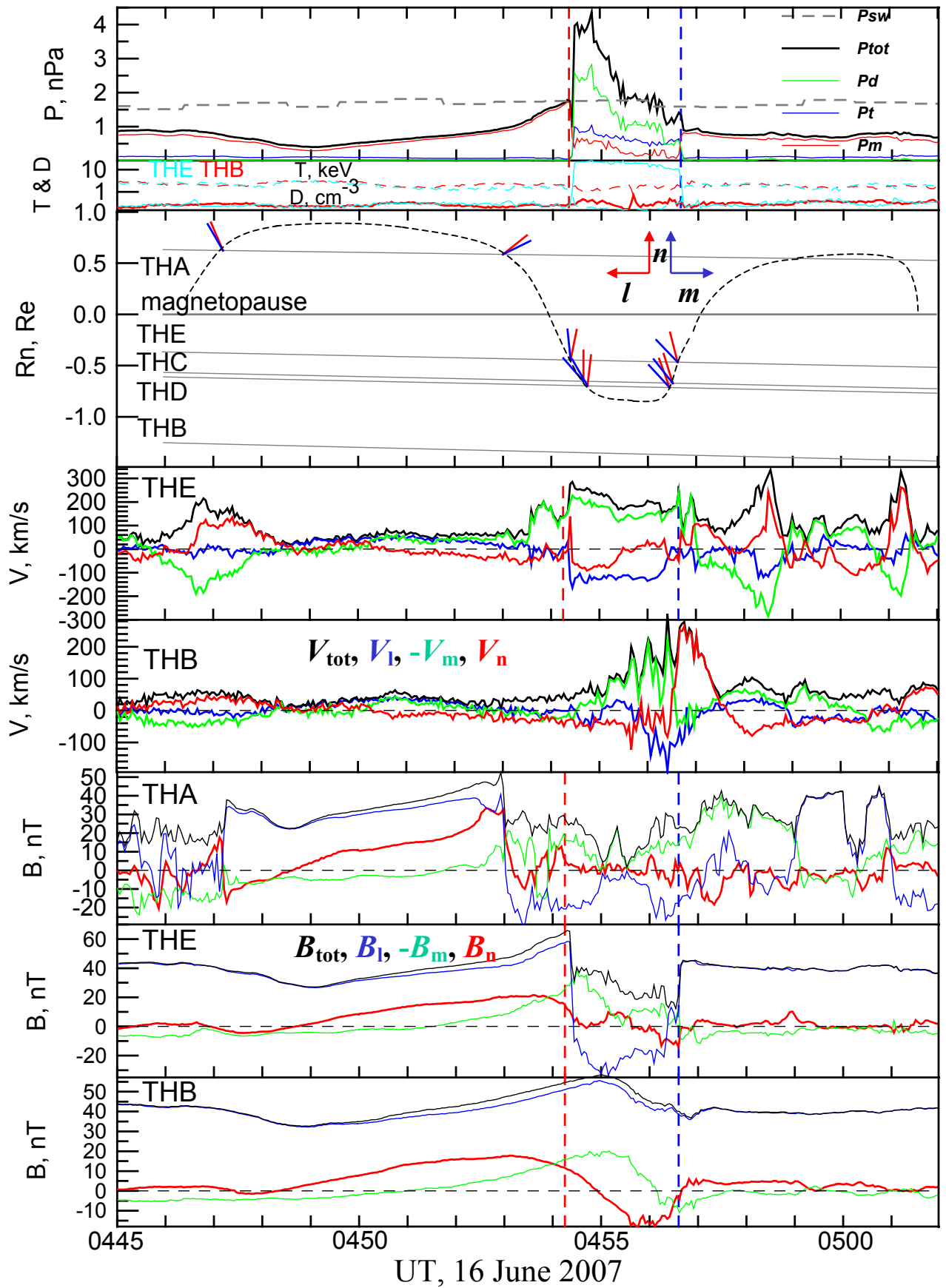

**Figure 2.** THEMIS observations of a magnetosheath plasma jet and magnetopause distortions in the frame of undisturbed nominal magnetopause by *Lin's et al.* [2010] at 0445 – 0502 UT on 16 June 2007. Top panel shows THE data on magnetic (red line), thermal (blue line), dynamic (green line) and total (black line) pressure in the magnetosphere and magnetosheath, and the total solar wind

pressure observed by ACE (gray dashed line). Vertical red and blue dashed lines depict, respectively, THE outbound and inbound magnetopause crossings caused by a magnetosheath plasma jet at 0444:10 to 0446:40 UT. The second panel shows plasma density (solid lines) and temperature (dashed lines) measured by THE (azure) and THB (red). The third panel shows orientation of the magnetopause normals in planes **l-n** (red segments) and **m-n** (blue segments). The normal distance of the THEMIS probes relative to the nominal magnetopause (gray thick horizontal line at $R_n=0$) is shown by thin gray lines. A sketch of the magnetopause location is shown by dashed line. Note that THB does not entrance to the magnetosheath. From the fourth to eighth panels show THEMIS observations of the magnetic field and plasma velocity in the nominal magnetopause coordinates. Total values and components **l**, **-m** and **n** are shown by black, blue, green and red curves, respectively. The jet causes a substantial distortion of the magnetopause shape and perturbations in the magnetic field and plasma fluxes.

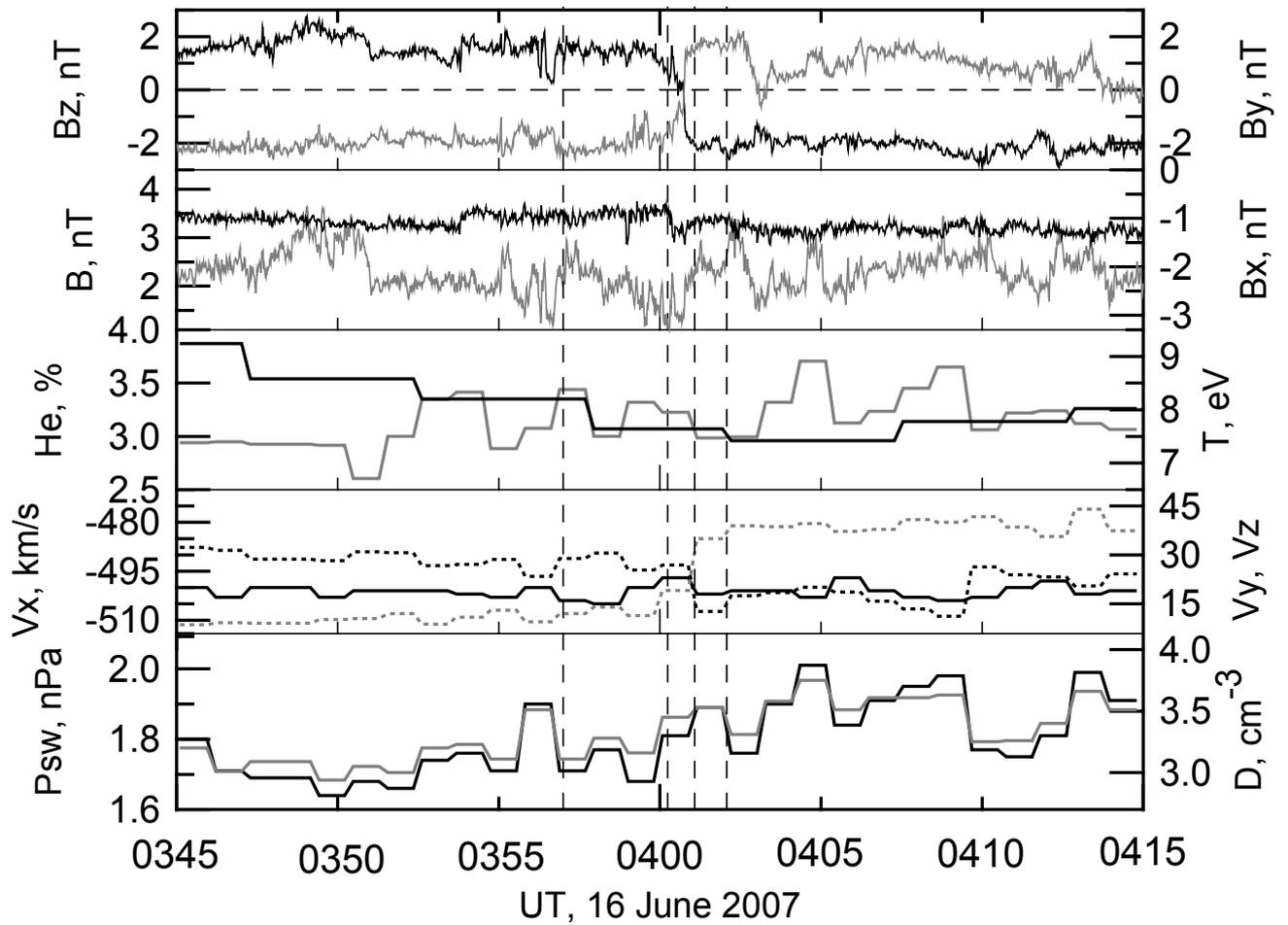

**Figure 3.** Upstream solar wind conditions measured by ACE on June 16, 2007 (from top to bottom): IMF components *B*z (black curve, left axis) and *B*y (gray curve, right axis) in GSM; IMF magnitude *B* (black curve, left axis) and *B*x component (gray curve, right axis); helium abundance (black curve, left axis) and proton temperature (gray curve, right axis); plasma velocity *Vx*-component (solid line, left axis), *Vy* and *Vz* components in GSM (black and gray dotted lines, respectively, right axis); sum of solar wind dynamic, thermal and magnetic pressures (black curve, left axis) and plasma density (gray curve, right axis). Vertical dashed lines restrict intervals used for analysis of a directional discontinuity (DD) centered at ~0400:46 UT.

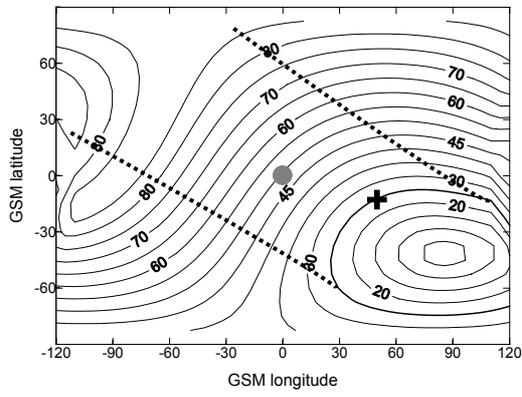

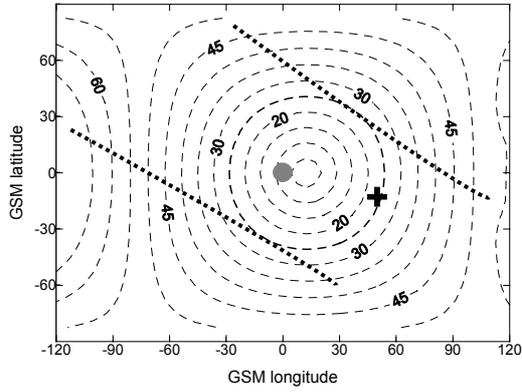

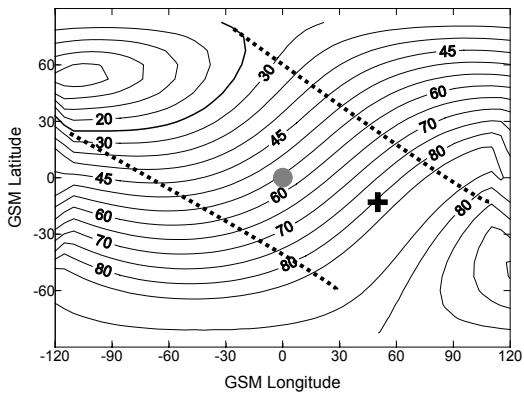

**Figure 4.** GSM maps of contours of equal cone angles $\theta_{Bn}$ (in degrees) between the IMF vector and the normal to *Chao's et al.* [2002] bow shock calculated for different orientations of the IMF observed by ACE: (a) at 0355 UT before the DD arrival; (b) at ~0400:46 UT right in the center of DD; (c) at 0406 UT, after the DD passing. The thick curves restrict regions of quasi-parallel bow shock ($\theta_{Bn} < 25°$). THEMIS location and subsolar point are indicated by the black cross and gray circle, respectively. Thick black dotted lines indicate a transitions corridor where the quasi-parallel bow shock and the foreshock transit from south-dusk to north-dawn region.

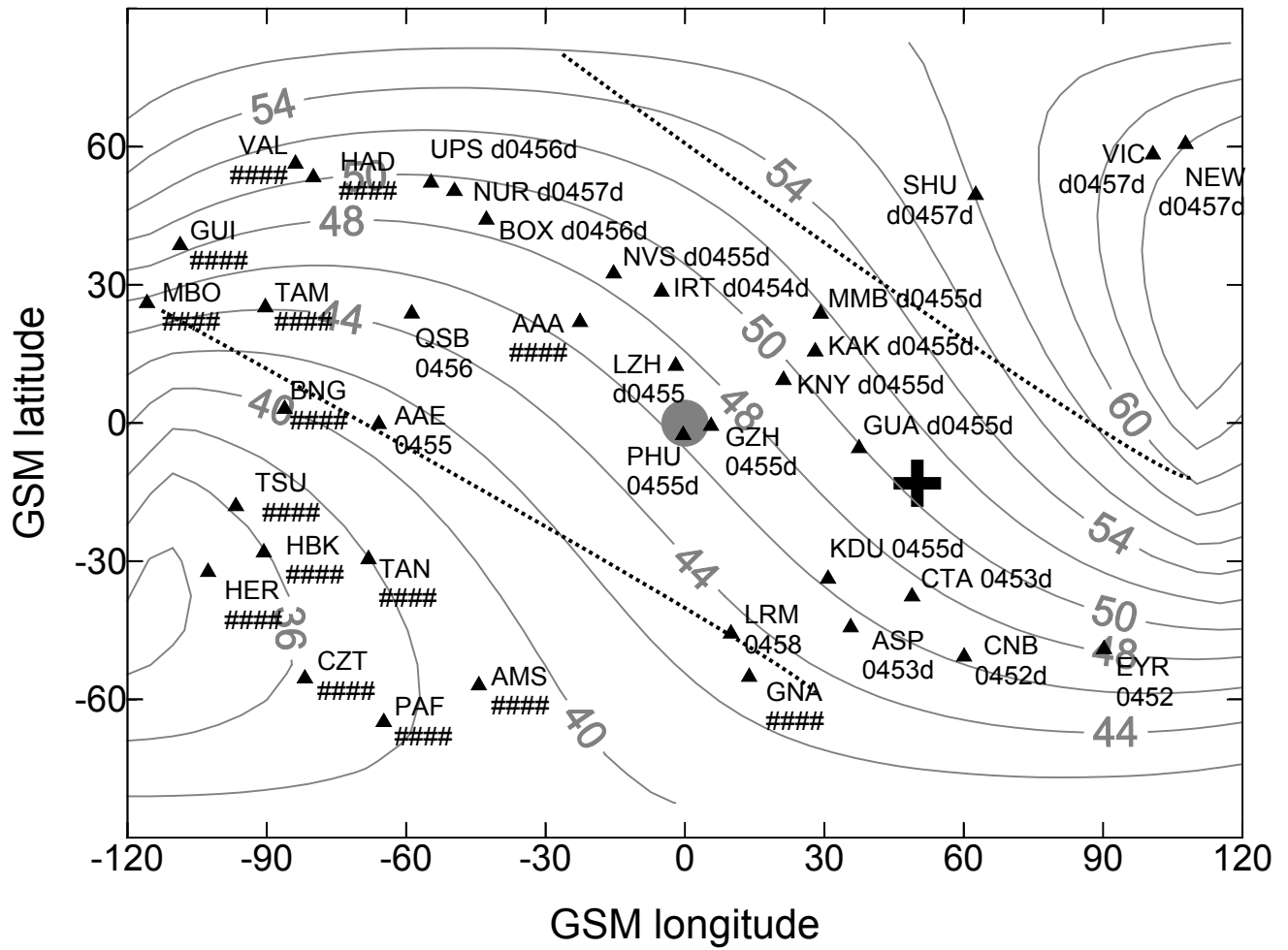

**Figure 5.** GSM map of INTERMAGNET stations (triangles) at 0455 UT on 16 June 2007. Four-digit numbers indicate time moment (UT) of the magnetic peak observation. Letter "d" before (after) the numbers indicates that the peak is preceded (followed) by a decrease of magnetic field. The absence of geomagnetic signatures is denoted by "####". Gray curves with numbers depict iso-contours of the DD propagation time from the ACE upstream monitor to the *Chao's et al.* [2002] bow shock. THEMIS location, subsolar point and the transition corridor are indicated by the black cross, gray circle and dotted lines, respectively. There are no prominent magnetic variations in the dawn sector. Magnetic pulse appears in the south-dusk sector and propagates toward north and dawn, mainly along the transition corridor. The magnetic variations with a pattern "decrease – peak – decrease" prevail in the northern hemisphere on the dayside and in the dusk sector.

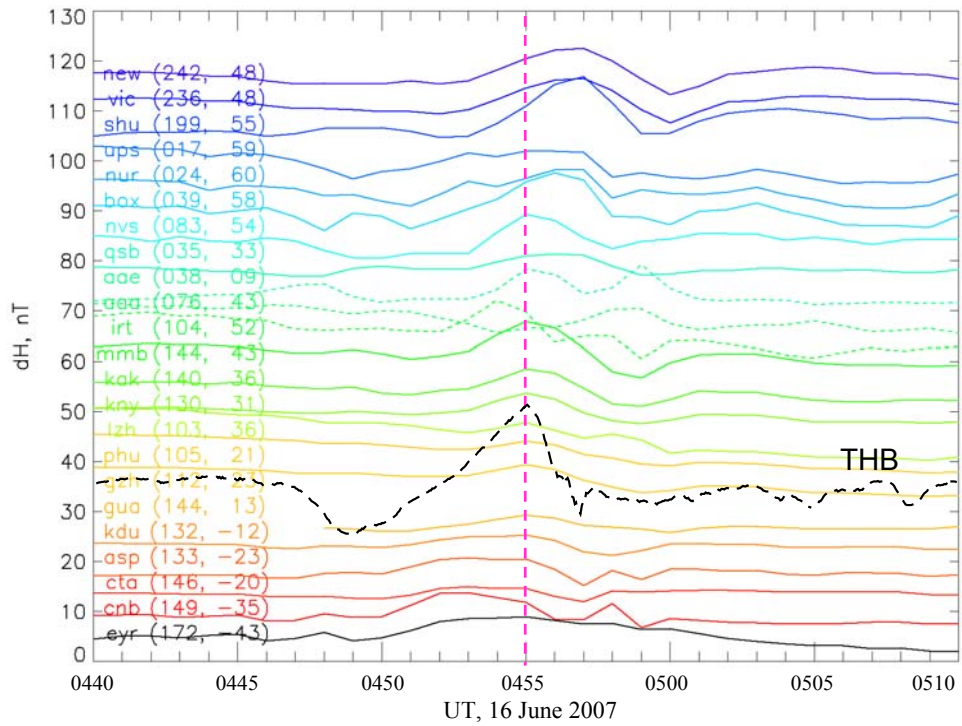

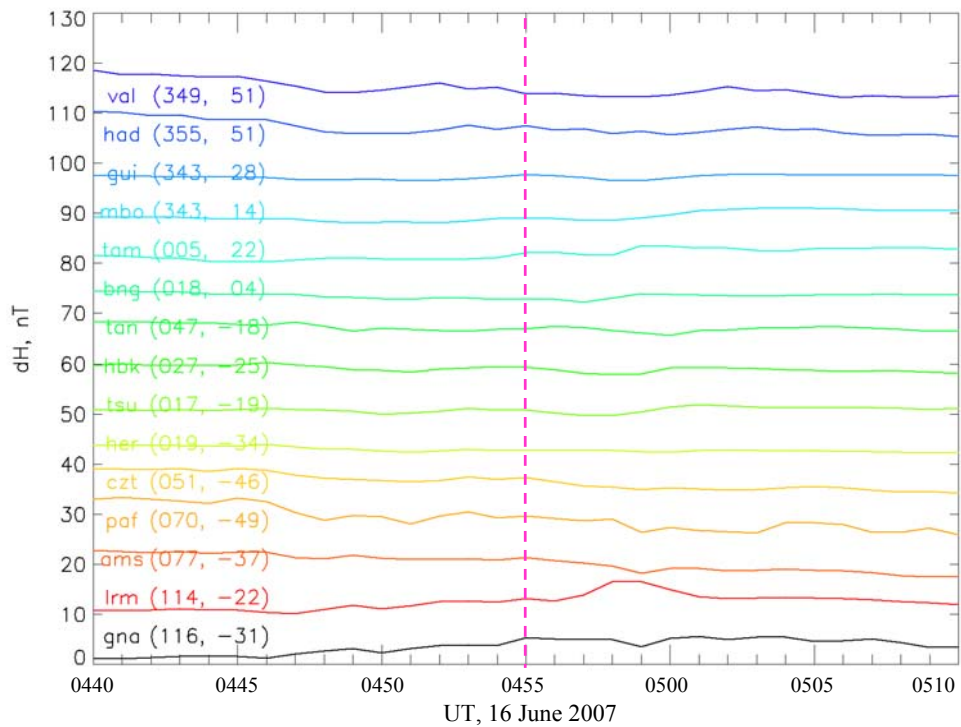

**Figure 6.** Variations of horizontal component d*H* of geomagnetic field: (a) prominent variations in the dayside and dusk sectors, THB magnetic field is shown for reference by black dashed curve, and (b) very weak variations in the dawn sector and in the beginning of DD interaction with the bow shock. Geographic longitude and latitude of the stations are indicated in parentheses. Vertical ink dashed lines depict the time 0455 UT of magnetic pulse maximum at THEMIS.